\documentstyle[aps,pre,epsf]{revtex}
\begin{document}
\title{\bf  Persistence  of   a  Continuous  Stochastic  Process  with
Discrete-Time Sampling: Non-Markov Processes}

\author{George C. M. A. Ehrhardt$^{(1)}$,  Alan J.  Bray$^{(1)}$,  
and Satya N. Majumdar$^{(2)}$}

\address{(1)Department of Physics and Astronomy, University of 
Manchester, Manchester, M13 9PL, UK \\ (2)Laboratoire  de Physique  
Quantique (UMR  C5626  du CNRS), Universit\'e Paul Sabatier, 31062  
Toulouse Cedex, France.}

\date{\today}

\maketitle

\begin{abstract}

\noindent  We  consider the  problem  of `discrete-time  persistence',
which  deals  with  the  zero-crossings  of  a  continuous  stochastic
process, $X(T)$, measured at  discrete times, $T=n\Delta\!  T$.  For a
Gaussian Stationary Process  the persistence (no crossing) probability
decays  as $\exp(-\theta_D  T)  = [\rho(a)]^n$  for  large $n$,  where
$a=\exp(-\Delta\!   T/2)$,  and  the  discrete  persistence  exponent,
$\theta_D$, is  given by  $\theta_D ={{\rm ln}  \rho \over 2  {\rm ln}
a}$.  Using  the ``Independent  Interval Approximation'', we  show how
$\theta_D$  varies  with $\Delta\!   T$  for  small  $\Delta\! T$  and
conclude  that  experimental measurements  of  persistence for  smooth
processes, such  as diffusion,  are less sensitive  to the  effects of
discrete sampling than measurements of a randomly accelerated particle
or  random  walker.  We  extend  the  matrix  method developed  by  us
previously [Phys.\  Rev.\ E {\bf  64}, 015151(R) (2001)]  to determine
$\rho(a)$ for  a two-dimensional  random walk and  the one-dimensional
random   acceleration   problem.   We   also   consider   `alternating
persistence', which corresponds to  $a<0$, and calculate $\rho(a)$ for
this case. 

\end{abstract}

\section{Introduction}

Persistence of  a continuous stochastic  variable has recently  been a
subject  of   considerable  interest  among   both  theoreticians  and
experimentalists.   Systems studied  include  {randomly driven  single
degrees                           of                           freedom
\cite{manyrefincmajumdarsdotxtpowalpha,burk,inelasticcollapseofarandomlyforcedparticle,inelasticcollapseofarandomlyforcedparticle2},
simple diffusion  from random initial  conditions\cite{IIA1,IIA2}, models of
phase
separation\cite{DBG,derrida1,derrida2,bray,satyaandsire2,majsire,crit,clement,krapiv,majcor,lee},
fluctuating interfaces\cite{krug,kallabis,tnd}, and reaction-diffusion
processes\cite{cardy,fisher,steve1,steve2}.   For   a  recent   review
see\cite{review}.   Persistence  is the  probability,  $P(t)$, that  a
fluctuating non-equilibrium  field, at  a particular space  point, has
not crossed  a certain threshold (usually  its mean value)  up to time
$t$.  In most systems studied, scale invariance implies that for large
$t$, $P$ exhibits a power-law decay, $P(t)\sim t^{-\theta}$, where the
persistence exponent $\theta$ is  non-trivial due to the dependence of
$P(t)$ on the whole history  of the system.  Experiments have recently
measured   $\theta$   for    the   coarsening   dynamics   of   breath
figures\cite{marcos},      liquid      crystals\cite{yurke},      soap
bubbles\cite{tam}, and diffusion of Xe gas in one dimension \cite{wong}.

In this paper  we consider the following problem:  in any experimental
(or  numerical)  measurement  of  $\theta$,  the  stochastic  variable
studied, $x(t)$, will have to  be sampled discretely.  It is therefore
possible  that $x(t)$  could  cross and  then  re-cross its  threshold
between samplings, resulting in a false-positive classification of the
persistence of  $x(t)$.  If the sampling is  logarithmically spaced in
time (as was the case  in \cite{wong}), then such undetected crossings
will  make  the  measured  persistence exponent  smaller  than  theory
predicts  (while if the  sampling is  uniform in  real time   only the
prefactor is changed). This problem has been studied in \cite{MBE} for
the  case of a  random walker  in one  dimension, a  simple Markovian,
`rough'   (i.e.\   with    a   fractal   distribution   of   crossings
\cite{fractalrefs}) process.  Here we extend  that work to  consider a
simple,  non-Markovian,  `smooth'   process:  a  randomly  accelerated
particle.   Since   most  of   the  more  complex   processes  studied
experimentally   are  `smooth',   this   paper  is   a  step   towards
understanding  the  effect discrete  sampling  has  on those  measured
persistence exponents.

There  is yet  another motivation  for studying  the persistence  of a
discrete sequence as opposed to that of a continuous process. It turns
out  that many  continuous processes  in nature  are  stationary under
translations  of time {\it  only} by  an integer  multiple of  a basic
period  (which  can  be  chosen  to  be  unity  without  any  loss  of
generality). For  example, the weather records have  this property due
to seasonal repetitions.  It has recently been shown\cite{MD} that for
a wide class of such processes, the continuous time persistence $P(t)$
is the  same as the  persistence $P(n)$ of the  corresponding discrete
sequence resulting from the measurement of the continuous data only at
discrete  integer   points.  In   general,  the  calculation   of  the
persistence  of a  discrete sequence  is much  harder than  that  of a
continuous  process,  except in  special  cases  where  $P(n)$ can  be
computed exactly\cite{MD,M}.  The tools  that have been developed over
the last decade  for studying the persistence of  a continuous process
are often not easily extendable  to the case of discrete sequences and
one needs  to invoke  new techniques, some  of which are  presented in
this paper.

The  layout of this  paper is  as follows:  In Section  II we  use the
Independent Interval Approximation  (IIA) \cite{IIA1,IIA2} to find the
first correction to $\theta$  for small $\Delta\!T$, where $\Delta\!T$
is the separation of samplings  in logarithmic time ($T=\ln t$).  This
gives us  some indication  of how significant  the effect  of discrete
sampling  is.  For  example,  for the  random  walk $\theta_D  -\theta
\propto -\sqrt{\Delta\!T}$  and so the  discrete exponent, $\theta_D$,
and its continuum  limit, $\theta$, begin to deviate  markedly as soon
as $\Delta\!T  > 0$.  For the random  acceleration problem,  one finds
$\theta_D  -\theta   \propto  -\Delta\!T$,  a   weaker  dependence  on
$\Delta\!T$,  while for  the  diffusion equation  from random  initial
conditions, $\theta_D  -\theta \propto -(\Delta\!T)^2$,  so the effect
of discrete sampling for small $\Delta\!T$ is much less significant in
this case.  These three  systems display an increasing `smoothness' of
the underlying process, a concept we will expand upon below.

In  Section   III  we  illustrate  our  general   method  for  solving
low-dimensional discrete persistence  problems by considering the case
of  a  random  walker  in one-dimension,  $\dot  x(t)=\eta(t)$,  where
$\eta(t)$ is  Gaussian white noise. A  brief account of  this work was
given  in \cite{MBE}.   We map  the problem  to a  Gaussian Stationary
Process  (GSP)  in  the  variable $X=x/\sqrt{t}$  by  transforming  to
logarithmic time. In the new variables, the random walk is represented
by an Ornstein-Uhlenbeck process. We  re-pose this problem in terms of
a backward Fokker-Plank Equation (BFPE)  and give the solution for the
continuum case where the persistence exponent has the well known value
$\theta=1/2$.  We  then formulate the discrete  persistence problem in
terms of  an eigenvalue integral  equation.  Employing a  power series
expansion of the integrand reduces  the problem to a matrix eigenvalue
equation, whose  largest eigenvalue gives us  the persistence exponent
measured   by   discrete  sampling.    The   concept  of   alternating
persistence, in  which the consecutive  measured values of $X$  lie on
alternate  sides  of the  threshold  value,  is  also introduced,  and
$\theta_D$ calculated for this case.

Having demonstrated the matrix method on a single variable problem, we
then apply it to two two-variable problems: the two-dimensional random
walk, $\dot  {\bf x}(t)={\bf \eta}(t)$,  in a wedge  geometry (Section
IV), and a randomly accelerated particle in one dimension, $\ddot x(t)
=\eta(t)$,  which  is a  simple  example  of  a non-Markovian  process
(Section  V).   Surprisingly,   for  alternating  persistence  in  the
continuum   limit  ($\Delta\!T=0$)   the  asymptotic   probability  of
surviving  one further  sampling,  $\rho$, is  non-zero, a  phenomenon
contrary  to our  earlier suggestion\cite{MBE}  that this  should only
occur for  `rough' processes.  Using scaled  variables and logarithmic
time,  the  random acceleration  problem  maps  onto  a damped  simple
harmonic oscillator, which we study  for the overdamped case using the
matrix method, finding $\theta_D$  as a function of $\Delta\!T$. Using
the results of a correlator  expansion developed in \cite{ctl} we also
study the  underdamped case,  an example for  which the  correlator is
oscillatory  (Section  VI).  We  show  that  when  the time  interval,
$\Delta\!T$,  between  measurements is  equal  to  the  period of  the
oscillation,  the  problem  is  identical  to  the  Ornstein-Uhlenbeck
process studied  in Section  III, while for  $\Delta\!T$ equal  to one
half  of  a  period  the   problem  reduces  to  that  of  alternating
persistence  in the  Ornstein-Uhlenbeck process.  The  paper concludes 
with a summary of the results.

\section{ Small $\Delta\!T$ Correction to $\theta$ }

The Independent Interval Approximation (IIA) \cite{IIA1,IIA2} uses the
assumption  that the  intervals between  zero-crossings of  a  GSP are
independently distributed.  Although this assumption is  not valid for
most  processes, it nevertheless  gives remarkably  accurate estimates
for $\theta$ in  many cases.  Here we use the  same assumption to find
the  first correction  to $\theta$  due  to discrete  sampling with  a
spacing in logarithmic time of $\Delta\!T$ for $\Delta\!T$ small.

As  $\Delta\!T$  is  increased  from  zero, the  first  correction  to
$\theta$ comes  from paths  which are always  positive apart  from one
undetected double  crossing between consecutive sample  times. Let the
probability of one such double crossing occurring in the interval $T =
n\Delta\!T$  be  $P_{dble}(n,\Delta\!T)$.  Then,  for  $T$ large,  the
probability, $P_D(T)$, that the stochastic variable is positive at all
$n$ samplings is given, to lowest order in $\Delta\!T$, by
\begin{equation} 
P_D(T) = P_0(T) +P_{dble}(n,\Delta\!T)
\label{005}
\end{equation} 
where $P_0(T)$  is the continuous-time  persistence probability, given
by    $P_0(T)   \sim    e^{-\theta    T}$   for    $T$   large,    and
$P_{dble}(n,\Delta\!T)$ is given by
\begin{equation} 
P_{dble}(n,\Delta\!T)   \propto   n   \,   e^{-\theta   n   \Delta\!T}
\int_0^{\Delta\!T} dT_1 \int_{T_1}^{\Delta\!T} dT_2 \, P_1(T_2-T_1)
\label{010}
\end{equation} 
where $P_1(T)$  is the probability distribution of  the interval size,
and  we have  assumed that  the durations  of different  intervals are
statistically independent. The latter assumption is precisely the IIA.
It  is  clear  from  Eq.\   (\ref{010})  that,  to  leading  order  in
$\Delta\!T$, we only require the form  of $P_1(T)$ in the limit $T \to
0$.  The  function $P_1(T)$  can be  found using the  IIA and  for the
processes  currently under  consideration --  the random  walk, random
acceleration  and  diffusion from  random  initial  conditions --  the
small-$T$  results  are  $P_1(T)  \propto 1/\sqrt{T}$,  $1$,  and  $T$
respectively.  All three cases  are incorporated  in the  general form
$P_1(T)\propto T^{\alpha}$, with  $\alpha=-1/2$, 0 and 1 respectively.
Using this form in Eq.\ (\ref{010}) gives
\begin{equation} 
P_D(T) = P_0(T)+\gamma \, n \, {\Delta\!T}^{\alpha+2} e^{-\theta T}
\label{020}
\end{equation} 
where  $\gamma$ is some  constant. Since  $T=n \Delta\!T$  and $P_0(T)
\sim e^{-\theta T}$, we have
\begin{equation} 
P_D(T) = A e^{-\theta T} (1+ B\,T\, {\Delta\!T}^{\alpha+1})
\label{030}
\end{equation} 
where $A$, $B$ are constants, and so, to lowest order in $\Delta\!T$,
\begin{equation} 
P_D(T) = A e^{-\theta T} e^{B\,T\, {\Delta\!T}^{\alpha+1}}.
\label{040}
\end{equation} 
Since $P_D(T) \sim e^{-\theta_D T}$, we obtain
\begin{equation} 
\theta_D = \theta -B \, {\Delta\!T}^{\alpha+1}.
\label{050}
\end{equation} 
For  the random walk,  random acceleration  and diffusion  from random
initial   conditions,  $\theta_D-\theta   \propto  -\sqrt{\Delta\!T}$,
$-\Delta\!T$, and $-(\Delta\!T)^2$  respectively.  From this we expect
that  the discrete  sampling is  important as  soon as  $\Delta\!T$ is
non-zero  for the random  walk, this  being related  to the  fact that
$P_1(T)  \to \infty$ for  $T \to  0$, i.e.\  that the  distribution of
crossings   is   fractal   and   hence   this   process   is   `rough'
\cite{fractalrefs}.    The  persistence   exponent   for  the   random
acceleration process is linear in $\Delta\!T$ and thus we expect it to
be less affected by the  discrete sampling for small $\Delta\!T$ .  In
sections III  and IV these  expectations will be confirmed  using the
matrix  method perturbative expansion  in $a=e^{-\Delta\!T  /2}$ about
$a=0$.   Finally, for  diffusion from  random initial  conditions, the
$(\Delta\!T)^2$  dependence indicates that  discrete sampling  will be
relatively unimportant  for small  $\Delta\!T$. Finally we  note that,
although Eq.\ (\ref{050})  has been derived using the  IIA, we expect
it to be  valid quite generally. In particular,  the $\alpha=2$ result
for     $P_1(T)$    has    been     proved    correct     by    Zeitak
\cite{zeitakshorttimeexpansion},  the  IIA  even  giving  the  correct
coefficient.

\section{ Random Walk in One Dimension }

Let us consider the simple case of a random walker in one dimension,
\begin{equation} 
\dot x(t)=\eta(t)
\label{10}
\end{equation} 
where  $\eta(t)$   is  Gaussian  white   noise  with  zero   mean  and
$\left<\eta(t)  \eta(t')\right>=2D \delta(t-t')$.  For  convenience we
map  this process,  which is  non-stationary  in time,  to a  Gaussian
Stationary   Process  (GSP)  by  defining new space and time variables 
$X = x/\sqrt{t}$, $T=\ln t$. Eq.\ (\ref{10}) then reads
\begin{equation} 
\frac{dX}{dT} =-\mu X(T) +\eta(T)\ ,
\label{20}
\end{equation} 
where $\left<\eta(T)\eta(T')\right>  = 2D \delta(T-T')$  and $\mu=1/2$
for  the  random  walker,  although  other  values  of  $\mu$  can  be
considered.   This  Ornstein-Uhlenbeck problem  can  be more  usefully
solved using the BFPE. Let $Q(X,T)$ be the probability that the random
walker, starting  from $X$  at time $T=0$,  does not cross  the origin
($X=0$) up  to time $T$.   Then from (\ref{20})  it can be  shown that
$Q(X,T)$ satisfies
\begin{equation} 
\partial  Q/\partial T  = D  \, {\partial}^2  Q/{\partial X}^2  -\mu X
\partial Q/\partial X\ ,
\label{30}
\end{equation}
with  initial   condition  $Q(X,0)=1$  for  all   $X>0$  and  boundary
conditions $Q(0,T)=0$ and $Q(\infty,T)=1$. The solution is
\begin{equation} 
Q(X,T)={\rm Erf}\left[{e^{-\mu T}\over {\sqrt{2 D' (1-e^{-2 \mu T})}}}
X\right]
\label{40}
\end{equation}
where ${\rm Erf}(x)$ is the error function and $D'=D/\mu $.  For large
$T$ (and positive $\mu $), $P(T) \sim X e^{-\mu T}$, which corresponds
to $P(t) \sim t^{-\theta}$ in real time, with $\theta = \mu$.

We now  consider discrete  persistence, i.e.\ the probability $Q_n(X)$
that  starting  at $X$  our  field/variable  is  positive at  all  the
discrete sample times  $T_1=\Delta \!  T$, $T_2=2 \Delta  \!  T$, ...,
$T_n=n \Delta \!  T$.  This  is relevant for experimental or numerical
determinations of $\theta$  since in practice one will  have to sample
only  at  discrete  points. In ref.\ \cite{wong}, for example,  the 
sampling is done logarithmically  in real time (which corresponds to 
sampling uniformly in logarithmic time).   
Note that $\theta_D$ will in general
be smaller than the continuum value $\theta$ since any even number of
crossings between samplings will go  unnoticed.  One can write down a
recurrence relation for $Q_n(X)$: 
\begin{equation} 
Q_{n+1}(X)=\int_0^{\infty} dY\, Q_n(Y) P(Y,\Delta \! T|X,0)
\end{equation}
where  $P(Y,\Delta  \!  T|X,0)$  is  the  Greens  function, i.e.\  the
probability that a particle starting at $X$ at time $0$ will be at $Y$
at time $\Delta \!  T$.  For a Gaussian process, $P(Y,\Delta \!T|X,0)$
can be found from the mean and variance of $X(T)$,
\begin{equation} 
P(Y,\Delta \! T|X,0) = {1\over \sqrt{2\pi D' (1-a^2)}}\,
\exp\left[-\frac{(Y-aX)^2}{2 D' (1-a^2)}\right]\ ,
\label{50}
\end{equation}
where $a=\exp(-\mu\Delta\!T)$. This gives us the discrete analogue of 
the BFPE (\ref{30}),
\begin{equation} 
Q_{n+1}(X)=\int_0^{\infty} dY\,Q_n(Y)  {1\over \sqrt{2\pi D' (1-a^2)}}
\,\exp\left[-\frac{(Y-aX)^2}{2 D' (1-a^2)}\right]\ ,
\label{60}
\end{equation}
with the continuum equation being recovered in the limit $\Delta\!T\to
0$.

Making the  change of variables  $x=X/\sqrt{D'(1-a^2)}$, $y=Y/\sqrt{D'
(1-a^2)}$, and $Q_n(X)=Q_n'(x)$ gives,
\begin{equation} 
Q'_{n+1}(x)={1\over   \sqrt{2    \pi}}   \int_0^{\infty}   dy\,Q'_n(y)
\exp[-(y-ax)^2/2]
\label{70}
\end{equation}
At late times, with $\mu  >0$, we expect $Q'_{n+1}(x)=\rho Q'_n(x)$ in
analogy  with  the  continuous  case  where  $P(T+\Delta  \!   T)=P(T)
e^{-\theta  \Delta \!   T}$, so  $\rho=e^{-\theta_D \Delta  \!T}$.  We
therefore  expect  that, for  large  $n$,  $Q_n(x)  \to \rho^n  q(x)$.
Substituting  this  into   (\ref{70})  gives  an  eigenvalue  integral
equation for $q(x)$
\begin{equation} 
\rho      \,      q(x)={1\over\sqrt{2\pi}}     \int_0^{\infty}dy\,q(y)
\exp[-(y-ax)^2/2]\ ,
\label{80}
\end{equation}
with an  eigenvalue $\rho(a)$ that  depends continuously on  $a$. Eq.\
(\ref{80}) has  an infinite number  of eigenvalues, but at  late times
only    the    largest   will    remain    since   $\sum_i    \rho_i^n
\approx\rho_{max}^n$  for  large $n$  (there  is  no other  eigenvalue
contiguous to the largest  eigenvalue).  By symmetrising the kernel in
(\ref{80}), one  can use  the variational method  to find  a rigorous
lower bound  for $\rho$ \cite{MBE}.  This method cannot be  applied to
the random  acceleration problem, however, since the  kernel cannot be
symmetrized.  For  $\Delta \! T \to \infty$,  correlations between the
$n$th and ($n$+2)th samplings are negligible, so
\begin{eqnarray} 
P_n & \approx & [{\rm Prob}\,({\rm two \ consecutive\ points\  have\ the\  
same\ sign})]^n \label{83} \\
  & = & \left({1\over 2}+{1\over 2}\left<{\rm  sign}[X(0)]\, {\rm
sign}[X(\Delta\!T)]\right>\right)^n \label{84} \\
 & = & \left({1\over 2}+{1\over 2}\,{2\over \pi} \arcsin[C(\Delta
\! T)]\right)^n\ , \label{85}
\end{eqnarray} 
where $C(T_2-T_1)=\exp[-\mu(T_2-T_1)]$ is the normalized autocorrelation 
function of $X(T)$, i.e.\ $C(T) = \langle X(T) X(0) \rangle/\langle X^2 
\rangle$.  Hence, for large $\Delta\!T$, 
\begin{equation} 
\rho ={1 \over 2} +{a \over \pi}+... \ \ .
\label{86}
\end{equation} 

Eq.\ (\ref{80}) can also be solved perturbatively by expanding the
$\exp(axy)$ term as a power series in $a$,
\begin{equation} 
\rho  \,q(x)={1\over  \sqrt{2   \pi}}  \int_0^{\infty}\,  dy\,  q(y)\,
e^{{-y}^2/2}\,e^{-a^2x^2/2}\,\sum_{m=0}^{\infty} {(axy)^m \over m!}\ .
\label{90}
\end{equation}
Defining
\begin{equation} b_m={a^{m/2} \over \sqrt m!} \int_0^{\infty}dy\,q(y) 
y^m e^{{-y}^2/2}
\label{100}
\end{equation}
gives
\begin{equation} 
\rho \,  q(x)={1\over \sqrt{2 \pi}}  e^{-a^2x^2/2} \sum_{m=0}^{\infty}
{b_m \over \sqrt{m!} }(\sqrt a x)^m\ .
\label{110}
\end{equation}
Multiplying through  by $e^{{-x}^2/2}x^n$ and integrating over $x>0$  
gives  
\begin{equation} \rho b_n = \sum_{m=0}^{\infty} A_{nm} b_m 
\label{120}
\end{equation}
where
\begin{equation} 
A_{nm}   ={1   \over   \sqrt{4   \pi  (1+a^2)}}   \left({2   a   \over
{1+a^2}}\right)^{(n+m)/2} {\Gamma[(n+m+1)/2] \over \sqrt {n! m!}}\ .
\label{130}
\end{equation}

By this  method, the eigenvalue integral equation (\ref{80}) has been
converted  to the  eigenvalue  matrix equation  (\ref{120}),  and  the
problem reduces to computing the largest eigenvalue of an $N \times N$
submatrix whose ($n$,$m$)th  elements decrease exponentially in $n+m$.
In\cite{MBE}  we   determined  $\rho$  numerically  to   one  part  in
$10^{12}$.  We have also  algebraically found $\rho_{max}$ as a series
expansion in $a$, the first four terms being:
\begin{equation} 
\rho={1 \over 2} +{a\over \pi} +{\pi-2 \over \pi^2} \, a^2 +{48-36 \pi
+7 \pi^2 \over 6 \pi^3} \, a^3 +\ldots .
\label{134}
\end{equation}

The coefficients  up to order $a^{49}$  are given in  appendix 2.  For
$\Delta \! T  \to 0$, $a \to 1$  and convergence becomes progressively
slower. However, the variational method still works in this region.

One may  also consider the  case of alternating persistence, i.e.\ the
probability that $X(n \Delta \! T)$ is positive for every even $n$ and
negative for every odd $n$ (or vice versa).  The limit of integration
in Eq.\ (\ref{80}) then changes, giving
\begin{equation} 
\rho \, q(x)={1\over \sqrt{2  \pi}} \int_{-\infty}^{0} dy \, q(y) {\rm
exp}[-(y-ax)^2/2]
\label{140}
\end{equation}
Substituting $y \to -y$, swapping  the limits of integration and using
$q(-y)=q(y)$ (since the process is symmetric around $y=0$) gives
\begin{equation} 
\rho  q(x)={1\over \sqrt{2 \pi}}  \int_{0}^{\infty} \,  dy \,  q(y) \,
{\rm exp}[-(y+ax)^2/2] \,
\label{150}
\end{equation}
which  is identical  to (\ref{80}),  but  with $a$  replaced by  $-a$.
Therefore  replacing $a$ by  $-a$ in  the matrix  equation (\ref{120})
will  give the  alternating persistence  eigenvalue  $\rho_{alt}$. The
case $|a|> 1$ may also be  considered, this corresponding to $\mu < 0$
and hence to an  unstable potential in the Ornstein-Uhlenbeck process.
For the alternating case,  $a<-1$, the calculation proceeds as before.
In   fact,    from   Eq.\   (\ref{130})   it   can    be   seen   that
$\rho(1/a)=\rho(a)|a|$  for  $a<1$,   so  one  need  only  investigate
alternating  persistence  in  the  range  $-1<a<0$. For  $a>1$  it  is
possible for the  walker to escape to infinity,  i.e.\ $q_n(x)\not \to
0$ for $n \to \infty$.  The asymptotic limit for $q(x)$ when $a>1$ may
be found using the matrix  method in the following way. Since $q(x)\to
1$ as  $x\to \infty$, it  is more convenient  to study a  new function
$u(x)$ defined by the relation,
\begin{equation}
q(x)=1- q(0)\int_x^{\infty} u(y) dy,
\label{151}
\end{equation}
where $q(0)$  is fixed by  Eq.\ (\ref{151}) with  $x=0$.  Substituting
$q(x)$ from Eq.\ (\ref{151}) into Eq.\ (\ref{80}) (with $\rho=1$ since
we are finding the stationary state) we find, after some algebra, that
$u(x)$ satisfies the integral equation
\begin{equation}
u(x)={a\over {\sqrt{2\pi}}}\left[e^{-{a^2}x^2/2}+
\int_0^{\infty}u(y)e^{-(y-ax)^2/2}dy\right]\ .
\label{int2}
\end{equation}
Note that, unlike Eq.\ (\ref{80}),  which determines $q(x)$ only up to
an   overall  multiplicative   constant,  Eq.\   (\ref{int2})   is  an
inhomogeneous equation  which fixes $u(x)$ absolutely,  as one expects
on physical grounds.

As before, we expand the factor $\exp(axy)$ in (\ref{int2}) as a power
series to obtain,
\begin{equation}
u(x) = \frac{a}{\sqrt{2\pi}}\,e^{-{a^2}x^2/2}\left[1 + \sum_{n=0}^\infty  
{a^n x^n \over {n!}}\int_0^\infty dy\, y^n u(y)
e^{-y^2/2} \right]\ .
\label{matrix11}
\end{equation}
Multiplying  through by  $x^m {a^{m/2}  \over  \sqrt{m!}} e^{-x^2/2}$,
integrating over positive $x$, and defining
\begin{equation}
c_n = \frac{a^{n/2}}{\sqrt{n!}}\int_0^\infty dy\,y^nu(y)\,e^{-y^2/2}\ ,
\label{matrix12}
\end{equation}
gives
\begin{equation}
c_n = a [A_{n0} +  \sum_{m=0}^{\infty} A_{nm} c_m]\ ,
\label{owt}
\end{equation}
where $A_{nm}$  is our previous matrix given by Eq.\ (\ref{130}), 
and 
\begin{equation}
u(x) = \frac{a}{\sqrt{2\pi}}\,e^{-a^2x^2/2}\,\left[1+\sum_{n=0}^\infty 
\frac{a^{n/2}x^n}{\sqrt{n!}}\,c_n\right]\ .
\label{u}
\end{equation}
Eq.\ (\ref{owt}) can be solved by matrix inversion:
\begin{equation}
c_n = a\sum_{m}(B^{-1})_{nm}A_{m0}\ ,
\label{owt2}
\end{equation} 
where
\begin{equation}
B_{nm} = \delta_{nm} -a A_{nm}\ .
\label{owt3}
\end{equation} 
The solution converges rapidly as a  function of the size, $N$, of the
matrix. In  practice, $N$  of order a  few hundred gives  very precise
results.  

Figure  \ref{f02}  shows $\rho(a)$  for  both  alternating and  normal
persistence, while  Figure \ref{f02pt5} shows  $\theta(a)$.  Note that
even  for $a=0.96$,  $\theta_D$ is  significantly below  the continuum
value of $1/2$. We argued in section II that the difference $\theta(1)
- \theta(a)$  decreases at  $\Delta\!T^{1/2}$, i.e.\  as $(1-a)^{1/2}$
for  $a  \to  1$, implying  a  square-root  cusp  at $a=1$  in  Figure 
\ref{f02pt5}.   Wong  et.al.\cite{wong},  measuring  persistence  in
one-dimensional diffusion  of Xe gas, sampled  logarithmically in time
such  that in  log-time their  $\Delta \!  T$ was  about 0.24.  If the
process  were a  random  walk, this  would  give $a  \sim  0.9$ and  a
difference  between $\theta_D$  and  the continuum  $\theta$ of  about
$20\%$.  For the  diffusion  equation that  describes the  experiment,
however,  the  approach  to  the  continuum is  more  rapid  (see  the
discussion in  section II), and  rather accurate results  are obtained
even for $a=0.24$ \cite{ctl}.

In Figure \ref{f02pt75} we show  the eigenfunction $q(x)$ for the case
$a=0.5$. This function was obtained by substituting the eigenvector of
the  matrix $A$  corresponding  to the  largest  eigenvalue into  Eq.\
(\ref{110}).   The asymptotic  behavior for  large $x$  is  $q(x) \sim
x^\nu$,  where $\nu  = \ln  \rho/\ln  a \simeq  0.530661$ for  $a=0.5$
\cite{MBE}  . Figure  \ref{f02pt75} also  contains a  plot  of $q(x)$,
given by Eq.\ (\ref{151}), for the case $a=2$, which corresponds to an
unstable potential.

\section{ Random Walk in 2-Dimensions: Wedge Geometry }

Having illustrated the perturbative  method on a simple Markovian case
for which  another approach (the variational method)  is available, we
intend to  study a simple  `smooth' non-Markovian process,  the random
acceleration problem, $\ddot x =\eta(t) $.  This process is equivalent
to  a Markovian  problem in  two variables,  $x(t)$ and  $v(t)$, where
$\dot v=\eta(t)$ and  $\dot x = v$. Before  dealing with this problem,
we will first consider another two-variable Markov process, namely the
random walk in two dimensions, $\dot x=\eta_x(t)$, $\dot y=\eta_y(t)$,
with  $\left<\eta_i(t)\right>=0$ and $\left<\eta_i(t)\eta_j(t')\right>
= 2D\delta_{ij}\delta(t-t')$.  Using the perturbative approach on this
pedagogical problem  will clarify its  use on the  random acceleration
problem.

Consider a wedge of angle $\alpha$ whose boundaries are absorbing, let
our random walker  start inside this wedge at  radial position $r$ and
angle  $\phi$, with  $0 \le  \phi \le  \alpha$. Making  the  change of
variable $R  = r/\sqrt{t}$, $T =  \ln t$, converts the  problem into a
GSP,   as  in  section   III.  The   corresponding  BFPE,   i.e.\  the
two-dimensional analog of Eq.\ (\ref{30}), is
\begin{equation} 
\partial Q/\partial T = {\nabla^2}Q - \mu R \partial Q/\partial R
\label{160}
\end{equation}
where $\mu=1/2$  for the  random walk, though  we will treat  $\mu$ as
arbitrary,  and  $Q(R,\phi,T)$  is  the survival  probability  of  the
particle at time $T$ given  that it started at $(R,\phi)$. The initial
condition is $Q(R,\phi,0)=1$ for  $R>0$ and $0<\phi<\alpha$, while the
boundary     conditions    are     $Q(R,0,T)=0$,    $Q(R,\alpha,T)=0$,
$Q(0,\phi,T)=0$,  and  $Q(\infty,\phi,T)=1$  for $0<\phi<\alpha$.  The
solution can be obtained using separation of variables. Its asymptotic
form for $T \to \infty$ at fixed $R$ and $\phi$ is
\begin{equation}
Q(R,\phi,T) \propto R^{\pi/\alpha}\sin(\pi\phi/\alpha)
\exp(-\mu\pi T/\alpha)\ ,
\end{equation}
giving $\theta=\mu \pi/\alpha$.
Now  consider  the discrete  persistence.  The  analog of  equation
(\ref{80}) is,
\begin{equation} 
\rho  \, q({\bf  r})  = {1  \over  2\pi} \int  d{\bf  r'} q({\bf  r'})
\exp[-({\bf r'}-a {\bf r})^2/2]\ ,
\label{180}
\end{equation}
where $a=\exp(-\mu\Delta\!T)$ as before, ${\bf r} = (1-a^2)^{-1/2}{\bf
R}$, and the integration is over the wedge. Note that for $\alpha=\pi$
one can  do the integration  over $x$ and recover  the one-dimensional
result. In polar coordinates, Eq.\ (\ref{180}) becomes:
\begin{equation} 
\rho q(r,\phi) = {1 \over 2\pi} \int_0^{\infty} r' dr' \int_0^{\alpha}
d{\phi'} q(r',\phi') \exp[-(r'^2+a^2r^2 -2 a r r' \cos(\phi-\phi'))/2]\ .
\label{190}
\end{equation}
As before,  we expand the  exponential term containing  the mixed 
terms,
\begin{equation} 
\exp[a  r   r'  \cos(\phi-\phi')]=\sum_{m',n'}  a^{m'+n'}(rr')^{m'+n'}
[\cos(\phi)  \cos(\phi')]^{m'}  [\sin(\phi)  \sin(\phi')]^{n'} /  [m'!
n'!]\ ,
\label{200}
\end{equation}
and define
\begin{equation} 
b_{m'n'}= \left[a^{(m'+n')/2}/\sqrt{m'!n'!}\right]
\int_0^{\alpha}d\phi'  \int_0^{\infty} dr'\,r'^{m'+n'+1} e^{-{r'}^2/2}
[\cos^{m'}(\phi') \sin^{n'}(\phi')]\, q(r',\phi')\ ,
\label{210}
\end{equation}
giving 
\begin{equation} 
\rho q(r,\phi) = [e^{-a^2r^2/2}/(2\pi)] \sum_{m',n'}
r^{m'+n'}a^{(m'+n')/2}[\cos^{m'}(\phi)\sin^{n'}(\phi)]
b_{m'n'}/\sqrt{m'! n'!}\ .
\label{220}
\end{equation}
Multiplying through by
\begin{equation} 
\left(a^{(m+n)/2}/\sqrt{m!n!}\right)r^{m+n+1}e^{-r^2/2}
\cos^m(\phi) \sin^n(\phi)\ ,
\label{224}
\end{equation} 
and integrating over the wedge, gives
\begin{equation} \rho \, b_{mn} = \sum A_{mn, m'n'} b_{m'n'}\ .
\label{230}
\end{equation}
Note that, by  regarding $mn$ as a single index,  we can treat $A_{mn,
m'n'}$ as  a conventional matrix when solving  numerically for $\rho$.
The matrix elements are,
\begin{equation} A_{mn, m'n'}=
   \,   {1\over  4   \pi  a   \sqrt{m!n!m'!n'!}}   \left[{2   a  \over
a^2+1}\right]^{(m+n+m'+n'+2)/2} \Gamma[(m+n+m'+n'+2)/2] J_{mn,m'n'}
\label{240}
\end{equation}
where
\begin{equation} 
J_{mn,m'n'}=\int_0^{\alpha}d\phi\,[\cos(\phi)]^{m+m'}[\sin(\phi)]^{n+n'}\ .
\label{250}
\end{equation}

The problem has  been reduced to an eigenvalue  matrix equation in four 
variables which we solve numerically.   We have also found $\rho$ as a
series  expansion  in $a$.   As  before  we  can consider  alternating
persistence,  i.e.\ the  probability  that between  each sampling  the
particle  alternates   between  $0<\phi<\alpha$  and  $\pi<\phi<\alpha
+\pi$.  Figure \ref{f10} shows a plot of the results. The case 
$\alpha = \pi$ corresponds to  the one-dimensional
case  while $\alpha=0$  and $\alpha=\pi$  trivially give
$\rho$=0 and $\rho$=1 respectively.

For $\Delta T  \to \infty$, $\rho \to \alpha /2  \pi$ since the system
re-equilibrates  between   samplings:  In  this   limit  the  function
$\rho(\alpha)$ is a straight line from (0,0) to ($2\pi$,1).

In the  limit $\alpha \to  0$, we can  expand the $\phi'$  integral in
(\ref{190})  to first  order in  $\alpha$, and  set  $\phi=0=\phi'$ to
obtain
\begin{equation} 
{\rho \over  \alpha} q(r,0)  = {1 \over  2\pi} \int_0^{\infty}  r' dr'
q(r',0) \exp[-(r'^2+a^2r^2 -2 a r r')/2]\ ,
\label{270}
\end{equation}
which  is  identical  to  the  one-dimensional  random  walk  equation
(\ref{80})  but  with  $r'dr'$  rather  than just  $dr'$,  and  $\rho$
replaced by $\sqrt{2 \pi} \rho/ \alpha$.  This gives a matrix equation
similar  to  that   for  the  $d=1$  case,  but   with  $\rho  \propto
\alpha$. Also $\rho$ is a  monotonically increasing function of $a$ as
in  the $d=1$  random walk  case.  The  same argument  applies  to the
alternating  case  (where  $a  \to  -a$).  This  explains  the  linear
dependence  of  $\rho$  on  $\alpha$  for  small  $\alpha$  in  Figure
\ref{f10}  and   the  qualitative  $a$-dependence  of   the  slope  at
$\alpha=0$.

In the  limit $\alpha \to  2\pi $,  we  expect $\rho  \to \alpha/2\pi$
since the ``mean free path'' of the particle between samplings is very
much  greater than  the width  of the  absorbing region,  so  that the
particle  does  not  ``see''   the  absorbing  region  and  hence  the
probability distribution is not affected by it .  Thus doubling $(2\pi
-\alpha)$ doubles the probability of being absorbed.  This is true for
any  $\Delta \!   T$ although  since the  ``mean free  path'' $\propto
\sqrt{\Delta \!  T}$, as $\alpha$  is decreased from $2\pi$ the linear
relationship  will fail  sooner for  smaller $\Delta  \!T$.   The same
argument applies  for the  alternating case, but  since there  are two
alternating absorbing  regions, the probability  distribution has more
time  to  equilibrate near  each  absorbing  region,  thus the  linear
dependence breaks down  at smaller $\alpha$ for a  given $\Delta \!T$.
This explains  the linear  dependence near $\alpha  =2 \pi$  in Figure
\ref{f10}, and its breakdown as $\alpha$ decreases.
 
In the  continuum limit ($a=1$) we  have the result  $\theta = \mu\pi/
\alpha$. Since $\rho =a^{\theta/\mu}$, for $a=1$ we get $\rho =1 $ for
all $\alpha$.  The $a>0$ plots in Figure \ref{f10} are tending to this
limit for $a \to 1$.  In the alternating continuum limit ($a=-1$), the
particle cannot  survive for  $\alpha < \pi  $, since the  two sectors
that the particle must occupy  alternately are disjoint. For $\alpha >
\pi $ these two sectors overlap,  and it is clear that the alternating
persistence  problem ($a=-1$)  is equivalent  to the  usual persistence
($a=1$) but with  $\alpha$ replaced by $\alpha -\pi$,  i.e.\ $\theta =
\pi\mu /(\alpha -\pi)$.  So $\rho = 0 $ for $\alpha < \pi $, and $\rho
=1$ for $\alpha  > \pi$, i.e.\ we get a step  function at $\alpha =\pi
$.  The $a<0$ plots are tending to this limit for $a \to -1$.

From equation (\ref{240}) it  can be seen that $\rho(1/a)=a^2 \rho(a)$
for $a<1$, which  enables one to construct the results  for $a$ in the
range $a <  -1$ from those in the range $-1<a<0$.  The extra factor of
$1/a$ compared  to the  $d=1$ case  is due to  the phase  space factor
$rdr$ rather than $dx$ in the eigenvalue integral equation.

\section{ Randomly Accelerated Particle }

We  now consider  the non-Markovian  random acceleration  problem.  We
first  recast the equation  $\ddot x=\eta(t)$  as the  two first-order
equations  $\dot  v=\eta(t)$  and  $\dot  x=v$. After  the  change  of
variables  $V=v/t^{3/2}$,  $X=x/t^{1/2}$,  $T=\ln t$  these  equations
become
\begin{eqnarray} 
 dV/dT  & = & - \alpha  V + \eta(T) \label{279} \\  
 dX/dT & =  & - \beta X + V 
\label{280}
\end{eqnarray}
with  $\alpha=1/2$, $\beta=3/2$ for  the random  acceleration problem,
the  process  being `smooth'  for  any  $\alpha$,  $\beta$. The  noise
correlator  is $\langle  \eta(T)\eta(T')\rangle =  2D\delta(T-T')$. By
eliminating $V$, Eqs.\ (\ref{279}), (\ref{280}) can be written as
\begin{equation} 
d^2X/dT^2 + (\alpha +\beta )dX/dT +\alpha \beta X = \eta(T)\ ,
\label{290}
\end{equation}
which is  the equation  for an overdamped  simple-harmonic oscillator,
$\alpha=\beta$ corresponding to  critical damping.  Note that equation
(\ref{290})  is   symmetric  under  $\alpha   \leftrightarrow  \beta$.
Letting $T \to T/(\alpha  +\beta )$ and $X \to X/(\alpha+\beta)^{3/2}$
gives,
\begin{equation} 
\ddot X + \dot X +{\alpha /\beta \over (1+\alpha/\beta)^2}X = \eta(T)
\label{300}
\end{equation}
showing  that $\theta=\alpha f(\alpha/\beta)$ and  so only
differing ratios of $\alpha$ to $\beta$ need be considered.

For  continuum  persistence,   Sinai\cite{sinai}  and  also  Burkhardt
\cite{burk} have  shown that $\theta=1/4$ for  the random acceleration
problem ($\alpha=1/2$,  $\beta=3/2$).  For ratios  other than $\beta=3
\alpha$, no  analytic solutions have  been found, except for  the case
$\alpha/\beta \ll 1$ which we give in this paper.

To derive our usual eigenvalue  integral equation, we need to find the
propagator  $P(X,V,T+\Delta\!T|Y,U,T)$, that  is,  the probability  of
going from ($Y$,$U$)  at time $T$ to ($X$,$V$)  at time $T+\Delta\!T$.
For  a  Gaussian  process,   finding  the  means  and  variances  will
completely specify the distribution.

Let $\tilde  X=X-<X> \, ,\,  \tilde V=V-<V>$ and define $A=e^{-\alpha
\Delta T}$, $B=e^{-\beta \Delta T}$.  Then the propagator is,
\begin{equation} 
P(X,V|Y,U)= {1\over  {2 \pi \sqrt{{\rm Det} M}}} {\exp  \left[{-{1\over 2}
{(\tilde X,\tilde V) M^{-1} (\tilde X,\tilde V) }} \right] }
\label{310}
\end{equation}
where  $$M=\pmatrix{\left<\tilde X^2\right>  &  \left<\tilde X  \tilde
V\right> \cr \left<\tilde X  \tilde V\right> & \left<\tilde V^2\right>
} $$ and
\begin{eqnarray} 
\tilde X&=& X-YB-{U\over {\beta -\alpha  } } (A-B) \\ \tilde V&=& V-UA
\\ \left<\tilde X^2 \right>&=&{1\over {\alpha \beta (\beta -\alpha )^2
(\beta  +\alpha)}}   (  (\beta  -\alpha  )^2   -\beta  (\beta  +\alpha
)A^2+4\alpha \beta  AB-\alpha (\beta  +\alpha )B^2) \\  \left<\tilde X
\tilde V \right>&=&{1\over {\alpha  (\beta -\alpha ) (\beta +\alpha )}
}( (\beta -\alpha ) -(\beta +\alpha )A^2 +2 \alpha AB) \\ \left<\tilde
V^2 \right>&=&{1\over {\alpha}} (1-A^2)
\label{320}
\end{eqnarray}
and the eigenvalue-integral equation is,
\begin{equation} 
\rho   P_{\infty}(X,V)=  \int_0^\infty  dY   \int_{-\infty}^\infty  dU
P_{\infty}(Y,U) P(X,V|Y,U)\ ,
\label{330}
\end{equation}
where we  are working in  the forward variable  $P_{\infty}(X,V)$, the
probability of finding the particle  at position $X$ with velocity $V$
given that it has been found  at positive X at all previous samplings.
Making the change  of variable, ${X \over\sqrt {2 {\rm  Det} M}}= x $,
${V \over\sqrt{2{\rm Det} M} }= v $  , ${Y \over \sqrt {2 {\rm Det} M}
}= y  $ , ${U  \over \sqrt {2  {\rm Det} M}  }= u $,  $P_\infty(X,V) =
f(x,v)$ gives:
\begin{equation}
\rho   f(x,v)  = {1\over\pi}\sqrt{{\rm Det} M}\int_0^\infty dy\,
\int_{-\infty}^\infty du f(y,u) e^{-(\tilde x^2 <\tilde V^2> -2 \tilde
x\tilde v <\tilde X\tilde V> +\tilde v^2<\tilde X^2>)}
\label{340}
\end{equation}
Substituting 
\begin{eqnarray}
\tilde  v=v -Au,  \nonumber \\  
\tilde x=x-y  B -u  {{A-B}\over {\beta -\alpha }}\ ,
\label{344}
\end{eqnarray}
into Eq.\ (\ref{340}), the exponent will contain all pair combinations
of  $x$,$v$,$y$,$u$.    This  kernel   is  not  symmetrisable   in  $x
\leftrightarrow y$  with $v \leftrightarrow u$,  hence the variational
method  is  not  applicable.   Expanding  in  the  four  mixed  terms,
$xy$,$xu$,$vy$,$vu$, and using the same  method as before, we obtain a
matrix equation of the form
\begin{equation}
\rho I_{c,d}={\bf M}_{c,d,e,f} I_{e,f}
\label{345}
\end{equation}
(see appendix 1).  As previously, we have found $\rho$ numerically and
also as a series expansion in $a=\exp(-\Delta \!T/2)$. The results for
$\rho(a)$  are presented  in Figure  \ref{f20}, and  the corresponding
results for  $\theta(a)$ in Figure \ref{f25}. The  coefficients of the
expansion  up  to order  25  are  given in  Appendix  2  for the  case
$\alpha=1/2$, $\beta=3/2$.   As for the  random walk case,  the series
have  not  yet  converged  for  $a  \to  1$.   Since  we  now  have  a
four-dimensional matrix,  the problem is  more apparent since  we have
only been able to reach  $O(a^{25})$.  We extend  the curves by use of
Pad\'e Approximants \cite{DombAndGreenVol3}, which relies on $\rho(a)$
being `smooth', and so  do not work for the random walk  for $a \to 1$
because   of   the  $\sqrt{\Delta\!T}$   cusp.    Also,  (see  Figure 
\ref{f20}), the Pad\'e method does  not work in the alternating case
for  large $\beta$  because  of the  sharp  downturn for  $a \to  -1$.
Nevertheless,  in the  remaining  cases it  significantly extends  the
valid range  for $\rho$,  and we use  this for plotting  the remaining
alternating  persistence cases.   However, since  $\theta={1  \over 2}
\rm{ln}(\rho)/\rm{ln}(a)$,   when  $a   \to   1$  the   $1/\rm{ln}(a)$
accentuates the  slight error  in $\rho$.  To  remedy this, we  add an
extra term to the Pad\'e  polynomial in the numerator (or denominator)
to enforce the constraint  $\rho(1)=1$. Figure \ref{f20} shows plots
of the eigenvalue against $\exp(-\Delta \!  T/2)$ for $\alpha=1/2$ and
various  values   of  $\beta$.   Also  plotted  are   the  alternating
persistence results.

Surprisingly, for the continuum limit of alternating persistence $\rho
\not\rightarrow  0$. Indeed,  $\rho \approx  0.108$ for  all $\alpha$,
$\beta$.   Previously, we had  thought\cite{MBE} that  non-zero $\rho$
could only occur  for a `rough' process (which  has an infinite number
of  crossings in finite  time).  The  reason for  these results  is as
follows.  For alternating  persistence, $X$  (and similarly  $V$) must
cross zero  inside a time $\Delta  \! T$ so  for $\Delta \! T  \to 0$,
$X,V$  must  both  tend  to  zero. As  a  result,  Eqs.\  (\ref{279}),
(\ref{280})   become  $\dot   V(T)=\eta(T)$,   $\dot  X(T)=V(T)$,   or
equivalently  $\ddot X=\eta(T)$, thus  removing the  $\alpha$, $\beta$
dependence. This latter equation is invariant under changes in $T$ and
hence $\Delta \!   T$, provided we rescale $X$.   It therefore gives a
non-zero $\rho$ even  for $\Delta \! T \to 0$.   This implies that any
process  whose  equation  is  (or  becomes,  for  $\Delta\!T  \to  0$)
time-scale invariant has $\rho_{alt}  \not=0$ for $\Delta \!  T\to 0$.
For example, the diffusion  equation from random initial conditions is
equivalent to  the $n \to \infty$  limit of the process  $d^n x/dt^n =
\eta (t)$ and will reduce to $d^n  X/dT^n = \eta (T)$ for $\Delta \! T
\to 0$.

We  can  compare  our  results  to  exact  results  for  two  limiting
cases.   For  $\Delta   \!T   \to  \infty$,   as   before  we   expect
$\rho={1\over2}+{1\over \pi} \arcsin[C(\Delta \!  T)]$, where for this
case the normalized correlation function is
\begin{equation} 
C(\Delta\!T)={{\beta  e^{-\alpha  \Delta  \!   T }  -\alpha  e^{-\beta
\Delta \!T}}\over {\beta -\alpha}}
\label{465}
\end{equation} 
Hence  $\rho  =1/2  +{1  \over  \pi}  {\beta  \over  \beta  -\alpha  }
e^{-\alpha  \Delta   \!T}  +  O(e^{-2\alpha   \Delta  \!T},  e^{-\beta
\Delta\!T})$  for $\beta  > \alpha$.  For $\alpha  /\beta \to  0$, the
normalized  correlator,  Eq.\  (\ref{465}),  becomes  $C(\Delta\!T)  =
\exp(-\alpha\Delta\!T)$,  the random  walk correlator.  In  fact, Eq.\
(\ref{290}) reduces  to the  random walk equation,  $\dot X  = -\alpha
X+\eta  (T)$, if  the limit  $\alpha \ll  \beta $  is taken  after the
change  of variables  $X =  \chi/\beta\sqrt{\alpha}$, $T=\tau/\alpha$,
when  it  becomes clear  that  the  inertial  term is  negligible  for
$\alpha/\beta \to  0$.  For $\alpha /\beta  \ll 1$, we  can also solve
for  $\theta$  in the  continuum  limit,  $\Delta\!T=0$, by  expanding
around the Markov process $\dot X=-\alpha X+\eta(T)$ to first order in
$\sqrt{\alpha/\beta}$ \cite{majsire,alanetal,satyaandsire2}.
The correlator $C(T)$ can be written as
\begin{equation} 
C(T)=e^{-\alpha  T}   +{\alpha  \over  \beta-\alpha}   (e^{-\alpha  T}
-e^{-\beta T})\ .
\label{470}
\end{equation} 
Using the standard perturbation expansion \cite{alanetal} for $\theta$ 
for a process with
correlator $C(T)=e^{-\alpha T} +\epsilon  a(T)$, with $\epsilon \ll 1$, 
\begin{equation} 
\theta=\alpha  \left(1-  {2   \alpha  \over  \pi}{\alpha  \over  \beta
-\alpha}  \int_0^{\infty}  dT {a(T)  \over  (1-e^{-2 \alpha  T})^{3/2}
}\right)+O(\epsilon^2), 
\label{480}
\end{equation}
gives 
\begin{equation} 
\theta=\alpha \left(1-{2\over  \sqrt \pi}{\alpha \over  \beta -\alpha}
{\Gamma  \left[{\beta   \over  2  \alpha}\right]}/\Gamma  \left[{\beta
-\alpha \over 2 \alpha}\right] \right)\ .
\label{490}
\end{equation}
To first order in $\sqrt{\alpha/\beta}$, therefore, 
\begin{equation} 
\theta=\alpha     \left(1-{\sqrt{2\over    \pi}}{\sqrt{\alpha    \over
\beta}}\right) + O\left({\alpha \over \beta}\right) .
\label{500}
\end{equation}
For    $\alpha=1/2$,     $\beta=24$,    this    gives    $\theta=0.442
+O(\alpha/\beta)$,   which   is   consistent   with   the   value   of
$\theta=0.429$ obtained from the constrained Pad\'e.

The series for $\rho(a)$ in powers of $a$ agree with those found using
the correlator expansion \cite{ctl} for the various values of $\alpha$
and $\beta$ tried.  This was used  as a powerful check of the accuracy
of the correlator expansion.

As  before, the eigenfunction  $P_{\infty}(X,V)$ can  be reconstructed
from  the eigenvector  corresponding to  the largest  eigenvalue using
equation   (\ref{380}).    The   result   is   displayed   in   Figure
\ref{ddxeigenfn} for the standard case $\alpha=1/2$, $\beta=3/2$.

\section{Underdamped Noisy Simple Harmonic Oscillator }

We  now consider  the underdamped  case of  the noisy  simple harmonic
oscillator.  The persistence exponent for  this case is not known even
in the continuum  limit. Furthermore the correlator of  the process is
oscillatory, which  to our  knowledge has not  yet been  studied. From
equation (\ref{290}) it can be seen that the complex values
\begin{eqnarray} 
\alpha=\gamma+i \omega \nonumber \\ 
\beta =\gamma-i \omega
\label{50010}
\end{eqnarray}
correspond to an underdamped oscillator. The corresponding equation 
of motion is 
\begin{equation} 
\ddot X + 2 \gamma \dot X +(\gamma^2+ \omega^2) X = \eta(T)
\label{50020}
\end{equation}
where the angular  frequency of oscillation is $\omega$  and the decay
rate  is  $\gamma$.   Substituting   $\alpha$  and  $\beta$  into  the
correlator, Eq.\ (\ref{465}), gives:
\begin{equation} 
C(T)=\exp(-\gamma   T)  \,   [\cos(\omega  T)+{\gamma   \over  \omega}
\sin(\omega T)]\ .
\label{50030}
\end{equation}

Since $\alpha$, $\beta$  are complex, we choose to  study this process
using the correlator expansion  about $\Delta\!T \to \infty$ which was
developed in  \cite{ctl}.  We  merely substitute the  correlator, Eq.\
(\ref{50030}),   into   our    14th   order   series   expansion   for
$\rho(\Delta\!T)$ and hence find  $\theta_D$. Note that for the random
walk and random acceleration problems above, 14th order corresponds to
order $a^{14}$.  As before, $\theta_D = \gamma f(\omega/\gamma)$ so we
choose   to    keep   $\gamma=1/2$   and    vary   $\omega$.    Figure
\ref{ddxunderdampedtheta} shows plots  of $\theta$ against $\Delta\!T$
for various  values of  $\omega$. Also shown  are the random  walk and
persistence   and  alternating   persistence   exponents,  which   are
numerically identical to the  underdamped SHO when $\omega \Delta\!T =
2m\pi$ and  $\omega\Delta\!T = (2m+1)  \pi$ respectively, with  $m$ an
integer.   Note  that,  as  before,  the  series  have  not  converged
sufficiently for small $\Delta\!T$.

The (at first sight surprising) identity of the  random walk and  
underdamped SHO persistence   exponents for  certain   values   of  
$\Delta\!T$   is interesting,  and  the  explanation  rather simple:  
Since  these  are
Gaussian  processes of zero  mean, their  correlators possess  all the
information about them.   When studying discrete-sampling persistence,
we   are  in  effect   studying  a   sequence  of   random  variables,
$X_1,X_2,X_3,...X_n$,  whose  correlator  we know  since
$\left < X_p X_q \right  > = C(|p-q| \Delta\!T)$.  Any processes which
have identical correlators for all $C(m \Delta\!T)$, $m$ integer, will
have  identical  discrete  persistence  exponents for  this  value  of
$\Delta\!T$.   From equation  (\ref{50030}) it  can be  seen  that for
$\omega \Delta\!T  = 2m\pi$ the  correlator reduces to  the random
walk correlator, $\exp(-\gamma \Delta\!T)$, hence the agreement of the
persistence  exponents.  For  $\omega  \Delta\!T =  (2m+1) \pi$,  $C(m
\Delta\!T)=\exp(-\gamma  \Delta\!T) (-1)^m$,  that is,  the correlator
alternates in sign.  This corresponds to alternating persistence since
$X(m \Delta\!T)$ has in effect changed sign for $m$ odd.

At  this  point we  make  a general  comment  about  the variation  of
discrete persistence  exponents with $\Delta\!T$.   Consider a process
for which  we know $\theta(\Delta\!T)$ for some  range of $\Delta\!T$,
e.g. from  using the matrix  method or correlator expansion  for small
$\Delta\!T$: If  we change $\Delta\!T$ to  $\Delta\!T/m$, $m$ integer,
we  will  sample  at  exactly  the  same times  as  before,  plus  the
intermediate times.  Hence some paths that were persistent before will
be  excluded, and  so  the asymptotic  survival  probability decreases.
Thus,
\begin{equation} 
\theta_D(\Delta\!T/m) \geq \theta_D(\Delta\!T).
\label{50040}
\end{equation}
This of course does not imply that $\theta_D(\Delta\!T)$ is monotonic.
Taking    the   limit    $m   \to    \infty$,   and    assuming   that
$\theta_D(\Delta\!T)$  is smooth  for $\Delta\!T  \to 0$,  we  get the
result that
\begin{equation} 
\theta \geq \theta_D(\Delta\!T)\ ,\ \ \ \forall \Delta\!T\ .
\label{50050}
\end{equation}
This provides a lower bound on $\theta$ for when our large $\Delta\!T$
expansions do not converge up to the continuum limit.  We can use this
to see that, for $\gamma=1/2$, $\theta \geq 1.22$, $2.60$, $4.49$, and
$7.19$ for $\omega=2$, $4$, $8$, and $16$ respectively.

\section{Conclusion}
In this paper  we have extended our earlier  treatment of the discrete
persistence exponent for the random walk with an absorbing boundary at
the  origin  \cite{MBE}  to   the  two-dimensional  random  walk  with
absorbing boundaries  on a wedge, and the  random acceleration process
with an absorbing boundary at  the origin. While the latter is perhaps
the  simplest  continuous-time  non-Markovian  process,  both  of  the
processes discussed in this work have the simplifying feature that they
can be written as  a Markov process in  two dimensions. We
have  shown   that,  in  both  processes,   the  discrete  persistence
probability after  $n$ measurements, $Q_n(x)$, for  a process starting
at  $x$, has  the asymptotic  form  $Q_n(x) \sim  \rho^n q(x)$,  where
$\rho$  and  $q(x)$  are  the  largest  eigenvalue  and  corresponding
eigenfunction of  a certain  eigenvalue integral equation.   They have
been evaluated  to high precision by converting  the integral equation
into  a  matrix  eigenvalue  problem,  from  which  high-order  series
expansions in powers of  $a = \exp(-\mu\Delta\!T)$ have been obtained,
where  $\Delta\!T$ is a  uniform measurement  interval for a  Gaussian 
stationary  process (GSP).  The  random walk  and random  acceleration
problems have  been mapped onto  GSPs in logarithmic time,  $T=\ln t$,
and for these (and  similar) processes the calculations presented here
apply to the case of measurements uniformly spaced in $T$.

The case of  alternating persistence, in which the  measured values of
the stochastic variable take positive and negative values alternately,
has  been discussed  using a  formal continuation  of $a$  to negative
values.  The  case  $a>1$  (or $a<-1$  for  alternating  persistence),
corresponds, for a GSP, to  motion in an unstable potential. For $a>1$
there is a non-zero probability, $q(x)$, that the process, starting at
$x>0$, is  never measured  to be  negative, and we  have shown  how to
calculate  it, with  explicit results  (see, e.g.,  Figure 3)  for the
Ornstein-Uhlenbeck process.

For the random acceleration process, the corresponding GSP is a noisy,
overdamped harmonic  oscillator, $\ddot X  + (\alpha + \beta)\dot  X +
\alpha\beta  X =  \eta(T)$, with  $\alpha=1/2$ and  $\beta=3/2$.  This
process  is clearly  of  interest  for other  values  of $\alpha$  and
$\beta$,  since even  the value  of  $\theta$ in  the continuum  limit
($\Delta\!T \to  0$) is not  know exactly except  for $\beta=3\alpha$.
We have obtained a perturbative  result for the continuum limit in the
limit $\alpha  \ll \beta$. We  have also investigated  the underdamped
case, corresponding  to complex $\alpha$  and $\beta$, and  shown that
for  discrete measurements  with  time  step $\Delta  T$  equal to  an
integer number of oscillation  periods, the persistence properties are
identical to  those of  a random walk,  while for an  odd half-integer
number one recovers the alternating persistence of a random walk.
 
The methods  presented here become increasingly unwieldy  as the order
of the stochastic differential  equation increases.  Recently, a power
series approach has  been developed in which the  eigenvalue $\rho$ is
expanded in  powers of  the correlator, $C(k\Delta\!T)$,  evaluated at
integer multiples,  $k$, of the  time step between  measurements, with
the  maximum value  of $k$  depending on  the order  of  the expansion
\cite{ctl}.  While  not as powerful  as the matrix method  for systems
described by  low-order stochastic differential  equations, the series
approach  has  the  advantage that  it  can  be  applied to  any  GSP,
including  those (such  as diffusion  from random  initial conditions)
which cannot be described by a differential equation of finite order.

The  `simple'  persistence  problem  discussed  here  deals  with  the
probability  of detecting no  zero crossing  in $n$  measurements. The
statistics of the  number of zero crossings, i.e.\  the probability to
observe  $m$  crossings in  $n$  measurements,  is  also of  interest.
Results obtained  by applying both the  methods of this  paper and the
series expansion  approach of ref.\  \cite{ctl} will be reported in a
separate publication. 

This work was supported by EPSRC (UK).

\appendix
\section{Calculation of random-acceleration matrix equation}

Let the $xv$ coefficient in the exponent be $a_{xv}$ etc, then:
\begin{eqnarray}
a_{xx}&=&   -\left<\tilde    V^2\right>   {}\nonumber   \\   a_{xv}&=&
2\left<\tilde X\tilde V\right>  {}\nonumber \\ a_{vv}&=& -\left<\tilde
X^2\right>  {}\nonumber  \\  a_{yy}&=&  -B^2  \left<\tilde  V^2\right>
{}\nonumber \\ a_{yu}&=&  -2B{{A-B}\over {\beta -\alpha}} \left<\tilde
V^2\right> +2AB \left<\tilde X\tilde V\right> {}\nonumber \\ a_{uu}&=&
-A^2     \left<\tilde     X^2\right>     -\left({{A-B}\over     {\beta
-\alpha}}\right)^2  \left<\tilde  V^2\right>  +2A  {{A-B}\over  {\beta
-\alpha}}  \left<\tilde  X\tilde  V\right>  {}\nonumber  \\  a_{xy}&=&
2B\left<\tilde  V^2\right>  {}\nonumber  \\ a_{vy}&=&  -2B\left<\tilde
X\tilde   V\right>  {}\nonumber   \\  a_{xu}&=&   2{{A-B}\over  {\beta
-\alpha}}  \left<\tilde V^2\right> -2A\left<\tilde  X\tilde V\right>{}
\nonumber \\ a_{vu}&=&  2A\left<\tilde X^2\right> -2{{A-B}\over {\beta
-\alpha}} \left<\tilde X\tilde V\right>{}\ . \nonumber \\
\label{350}
\end{eqnarray}
Expanding the  exponent of Eq.\  (\ref{340}) in powers of  terms which
mix ($x$,$v$) with ($y$,$u$), we get,
\begin{eqnarray} 
\rho  f(x,v)  =  {1\over   \pi}\sqrt{{\rm  Det}  M}  \int_0^\infty  dy
\int_{-\infty}^\infty   du   f(y,u)   \sum_{n=0}^\infty  {1\over   n!}
(a_{xy}xy+a_{vy}vy  +a_{xu}xu+a_{vu}vu)^n  \nonumber  \\  \hspace{5cm}
\times \exp[-(a_{xx}x^2 +a_{xv}xv +a_{vv}v^2)]\ .
\label{360}
\end{eqnarray}
Letting
\begin{equation} 
I_{c,d}=\int_0^\infty  dx \int_{-\infty}^\infty  dv\,  f(x,v) x^c  v^d
\exp[-(a_{xx}x^2+a_{xv}xv+a_{vv}v^2 )]
\label{370}
\end{equation} 
gives
\begin{eqnarray} 
\rho  f(x,v)  =  {\sqrt{{\rm  Det}  M}  \over  \pi}  \sum_{e=0}^\infty
\sum_{f=0}^\infty{I_{ef}\over  (e+f)!}   {e+f\choose  e}  \sum_{r=0}^e
\sum_{s=0}^f   {e\choose  r}   {f\choose   s}  a_{xy}^{e-r}   a_{vy}^r
a_{xu}^{f-s}  a_{vu}^s x^{e-r+f-s}  v^{r+s} \nonumber  \\ \hspace{5cm}
\times \exp[-(a_{xx}x^2 +a_{xv}xv +a_{vv}v^2)]\ .
\label{380}
\end{eqnarray}
Multiplying through by $x^c v^d$,  and integrating over $x>0$ and over
all $v$, gives,
\begin{equation} 
\rho I_{c,d}={1\over \pi}\sqrt{{\rm Det} M} \, G_{c,d,e,f} \, I_{e,f}
\label{390}
\end{equation}
where
\begin{equation} 
G_{c,d,e,f}={1\over    (e+f)!}     {{e+f}\choose    e}    \sum_{r=0}^e
\sum_{s=0}^f  {e  \choose r}  {f  \choose  s} a_{xy}^{e-r}  a_{vy}^{r}
a_{xu}^{f-s} a_{vu}^{s} D_{e-r+f-s+c,r+s+d}
\label{400}
\end{equation}
and
\begin{equation} 
D_{a,b}=\int_0^\infty  dx\int_{-\infty}^\infty dv\,x^av^b \exp[-({\cal
A}v^2-{\cal B}xv+{\cal C}x^2 )]\ ,
\label{410}
\end{equation}
giving
\begin{equation} 
D_{a,b}={\sqrt  {\pi  \over  2}} \sum_{t=0}^{[b\over  2]}  {{b}\choose
b-2t} (2  {\cal A})^{-{(2t+1)\over  2}} \left({{\cal B}\over  {2 {\cal
A}}} \right)^{b-2t}  \left({\cal C} -{{{\cal B}^2}\over  {4 {\cal A}}}
\right)^{-\left({{a+b-2t+1}\over   2}\right)}   (2t-1)!!   \,   \Gamma
\left({{a+b-2t+1}\over 2} \right)\ ,
\label{420}
\end{equation}
where $[\frac{b}{2}]$ indicates the integer part of $\frac{b}{2}$ and
\begin{eqnarray} 
{\cal A}&=& -a_{vv} -a_{uu} \\
\label{430}
{\cal B}&=& \,\,\,\, a_{xv} +a_{yu} \\
\label{440}
{\cal C}&=& -a_{xx} -a_{yy}\ .
\label{450}
\end{eqnarray}

Thus the problem has been  reduced to computing the largest eigenvalue
of  an $N\times N\times  N\times N$  operator. As  in the  wedge case,
$G_{c,d,e,f}$  decouples   into  $G_{(c,d),(e,f)}$.   For  alternating
persistence,  the range of  integration over  $y$ in  Eq.\ (\ref{360})
should be from $-\infty$ to 0. Substituting $y \to -y$ and $u \to -u$,
and changing the limits of integration gives
\begin{eqnarray} 
\rho   f(x,v)   =   {\sqrt{Det   M}  \over   \pi}   \int_0^\infty   dy
\int_{-\infty}^\infty \! du  f(-y,\! -u) \sum_{n=0}^\infty {1\over n!}
(-a_{xy}xy -\! a_{vy}vy -\!  a_{xu}xu -\!  a_{vu}vu)^n \nonumber \\
\hspace{5cm} \times \exp[-(a_{xx}x^2 +a_{xv}xv +a_{vv}v^2)]\ .
\label{460}
\end{eqnarray}
Due to  the symmetry of the  system, $f(y,u) =  f(-y,-u)$, so changing
$a_{ij} $  $\to $  $-a_{ij}$ ($i=x,v$, $j=y,u$)  in the  matrix method
gives alternating persistence.

\section{Series for {\large $\rho$} for Random Walk and Random 
Acceleration.}

The coefficients for the random walk are:

$\rho(a) = 0.500000+ 0.318310 \,a^1  + 0.115668 \,a^2 + 0.021446 \,a^3
+ 0.015651  \,a^4 + 0.015762  \,a^5+ 0.000050 \,a^6 -0.003320  \,a^7 +
0.007597 \,a^8 + 0.007587  \,a^9 -0.004372 \,a^{10} -0.005964 \,a^{11}
+0.004840  \,a^{12}+ 0.007913  \,a^{13}  -0.001290 \,a^{14}  -0.006184
\,a^{15}  -0.000369  \,a^{16}+  0.004205 \,a^{17}+  0.001467  \,a^{18}
-0.000757  \,a^{19}+ 0.000989  \,a^{20}+  0.000385 \,a^{2}1  -0.003272
\,a^{22}  -0.002408  \,a^{23}+  0.003354 \,a^{24}+  0.004972  \,a^{25}
-0.000072  \,a^{26} -0.003737  \,a^{27}  -0.001786 \,a^{28}+  0.000650
\,a^{29}+  0.000270 \,a^{30}+  0.000415  \,a^{31}+ 0.002188  \,a^{32}+
0.001629  \,a^{33}  -0.001600  \,a^{34} -0.002448  \,a^{35}+  0.000213
\,a^{36}+  0.001451 \,a^{37}  -0.000501  \,a^{38} -0.001372  \,a^{39}+
0.000740  \,a^{40}+  0.002125  \,a^{41}+ 0.000743  \,a^{42}  -0.000602
\,a^{43}  -0.000379  \,a^{44}  -0.000732 \,a^{45}  -0.001956  \,a^{46}
-0.001143 \,a^{47} +0.001780 \,a^{48} +0.002824 \,a^{49}+ O(a^{50}).$

The coefficients for the random acceleration are:

$\rho(a) = 0.500000 + 0.477464 \,  a + 0.021519 \, a^{2} + 0.000314 \,
a^{3} +  0.035886 \, a^{4} - 0.063298  \, a^{5} + 0.029548  \, a^{6} +
0.032884 \, a^{7}  - 0.061472 \, a^{8} + 0.020790  \, a^{9} + 0.030977
\, a^{10}  - 0.039237  \, a^{11}  + 0.016107 \,  a^{12} +  0.012402 \,
a^{13} - 0.035066 \, a^{14} +  0.021396 \, a^{15} + 0.020271 \, a^{16}
- 0.027422  \, a^{17}  + 0.002006  \, a^{18}  + 0.005649  \,  a^{19} -
0.009557  \,  a^{20} +  0.018307  \, a^{21}  +  0.002099  \, a^{22}  -
0.023960\, a^{23}+ 0.007071\, a^{24} + 0.012405\, a^{25} + O(a^{26}).$

The coefficients  in both  series have been  truncated to  six decimal
places for  brevity. In the  plots presented in the  paper, sufficient
precision has been retained in the coefficients to ensure the accuracy
of the plots and of any quoted values for $\rho$ and $\theta$.

\newpage

\begin{figure}
\epsfbox{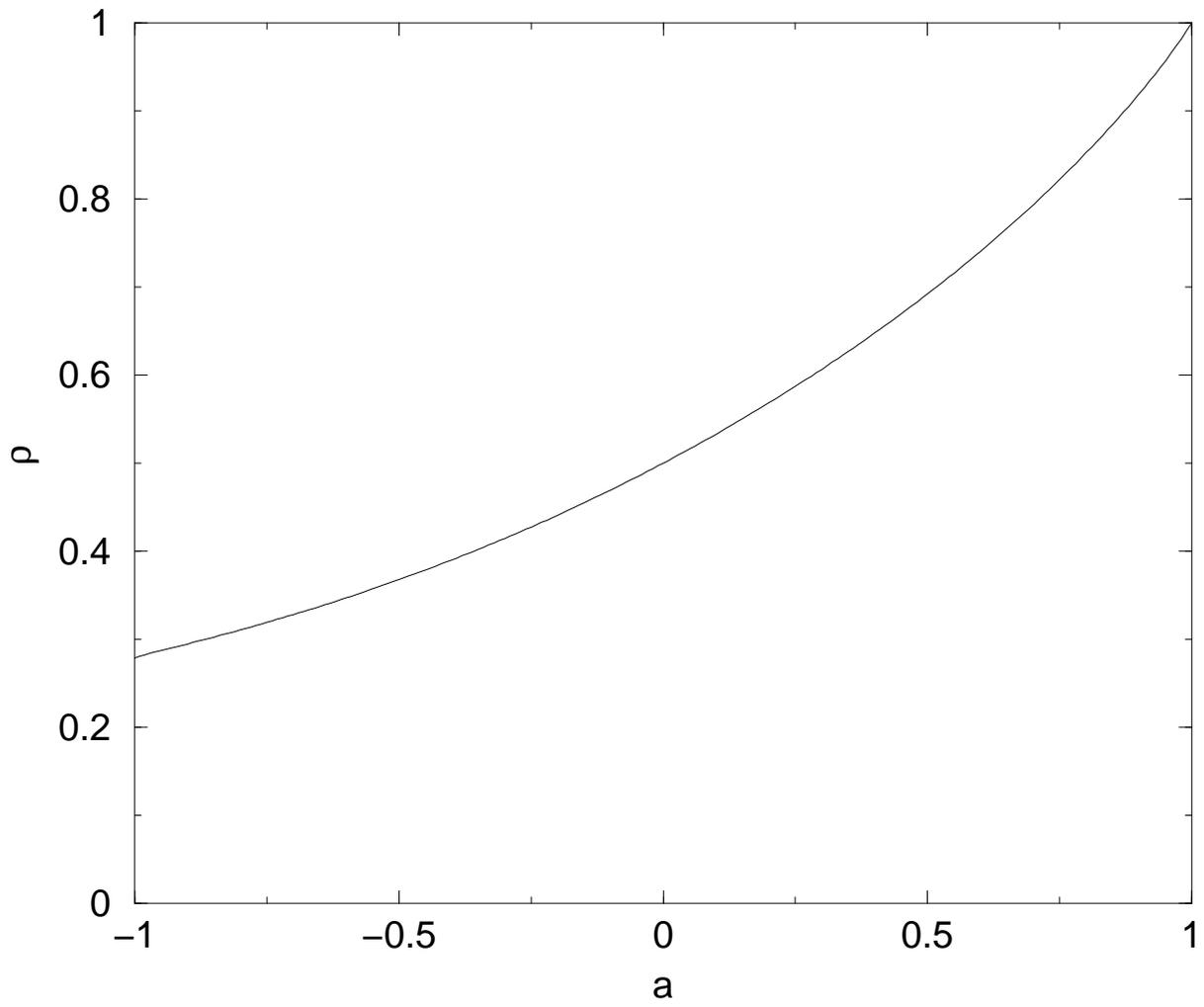}
\caption{Plot  of $\rho  (a)$ vs  $a$ ($a=e^{-\mu  \Delta\!T}$), where
$a<0$ corresponds to alternating persistence}
\label{f02}
\end{figure}

\begin{figure}
\epsfbox{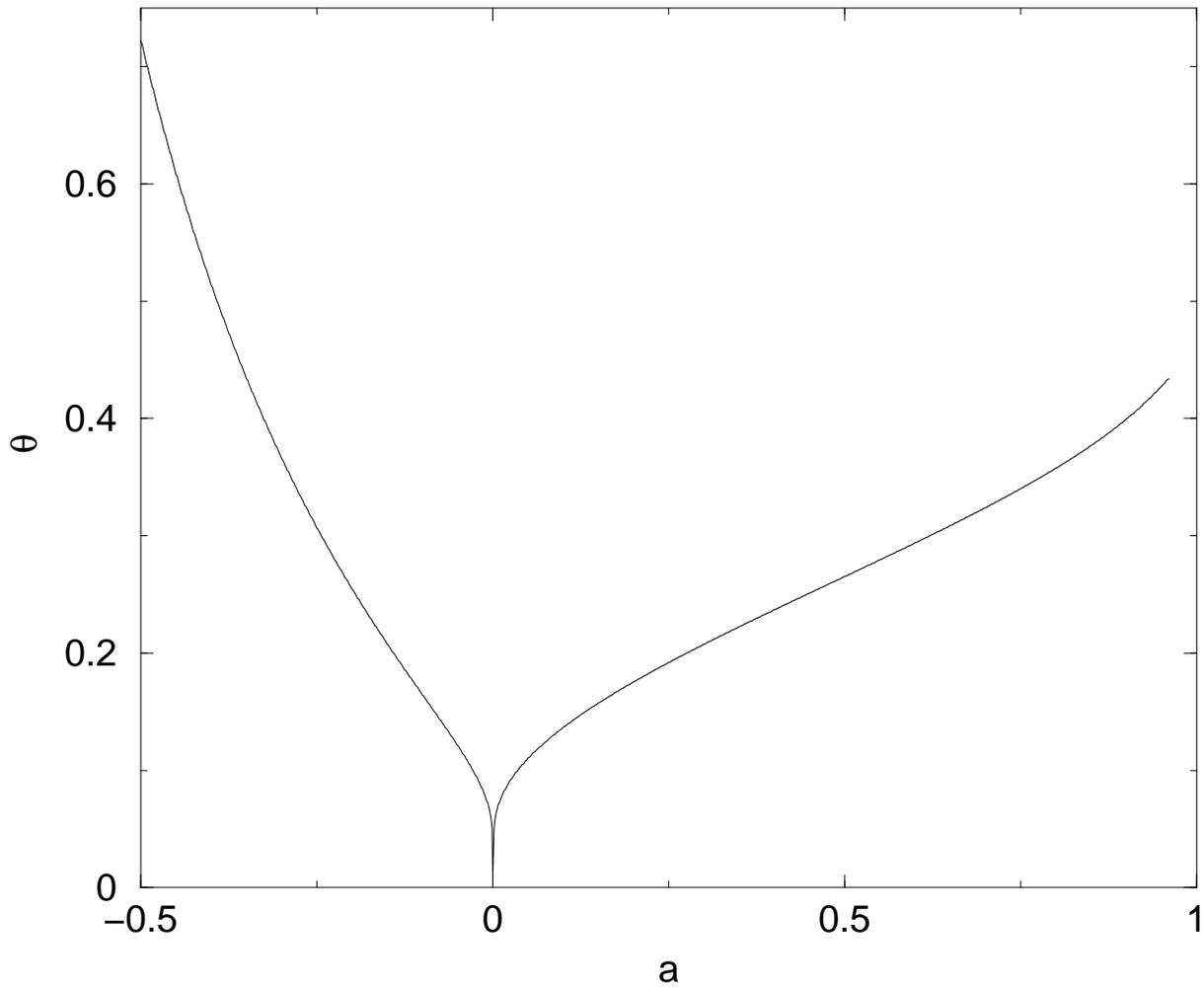}
\caption{Plot of  $\theta (a)$ vs $a$  ($a=e^{-\mu \Delta\!T}$), where
$\theta=\ln[\rho(a)]/(2 \ln|a|)$ and  $a<0$ corresponds to alternating
persistence.  For $a  \to 1$, the series has not  yet converged as the
$1/\ln(a)$  term amplifies  the  small numerical  error in  $\rho(a)$,
whilst for $a \to -1$, $\theta \sim 1/\ln(|a|)$.}
\label{f02pt5}
\end{figure}

\begin{figure}
\epsfbox{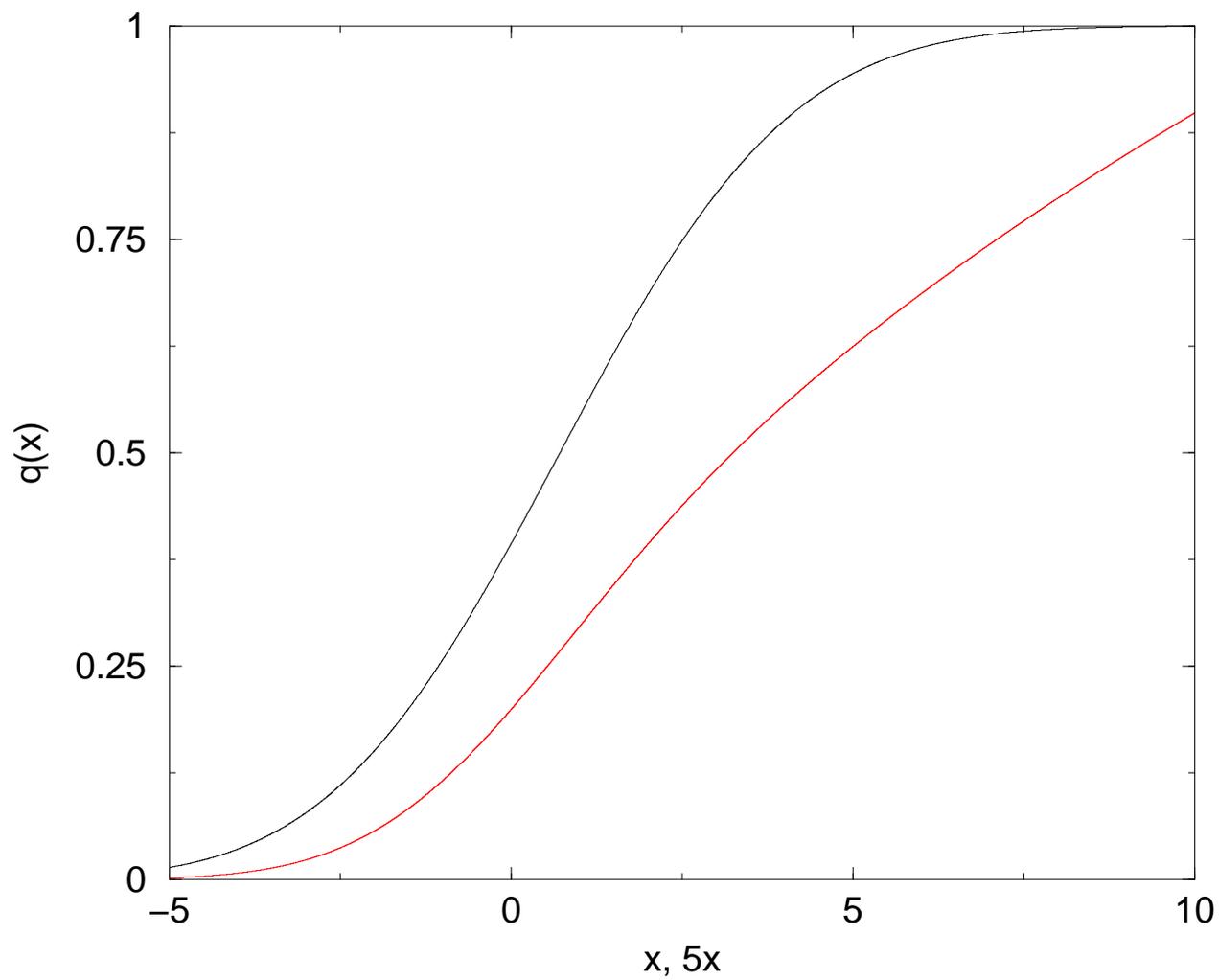}
\caption{Plot  of eigenfunctions  q(x) for  $a=0.5$ (lower  curve) and
$a=2.0$ (upper curve, abscissa=$5x$)}
\label{f02pt75}
\end{figure}

\begin{figure}
\epsfbox{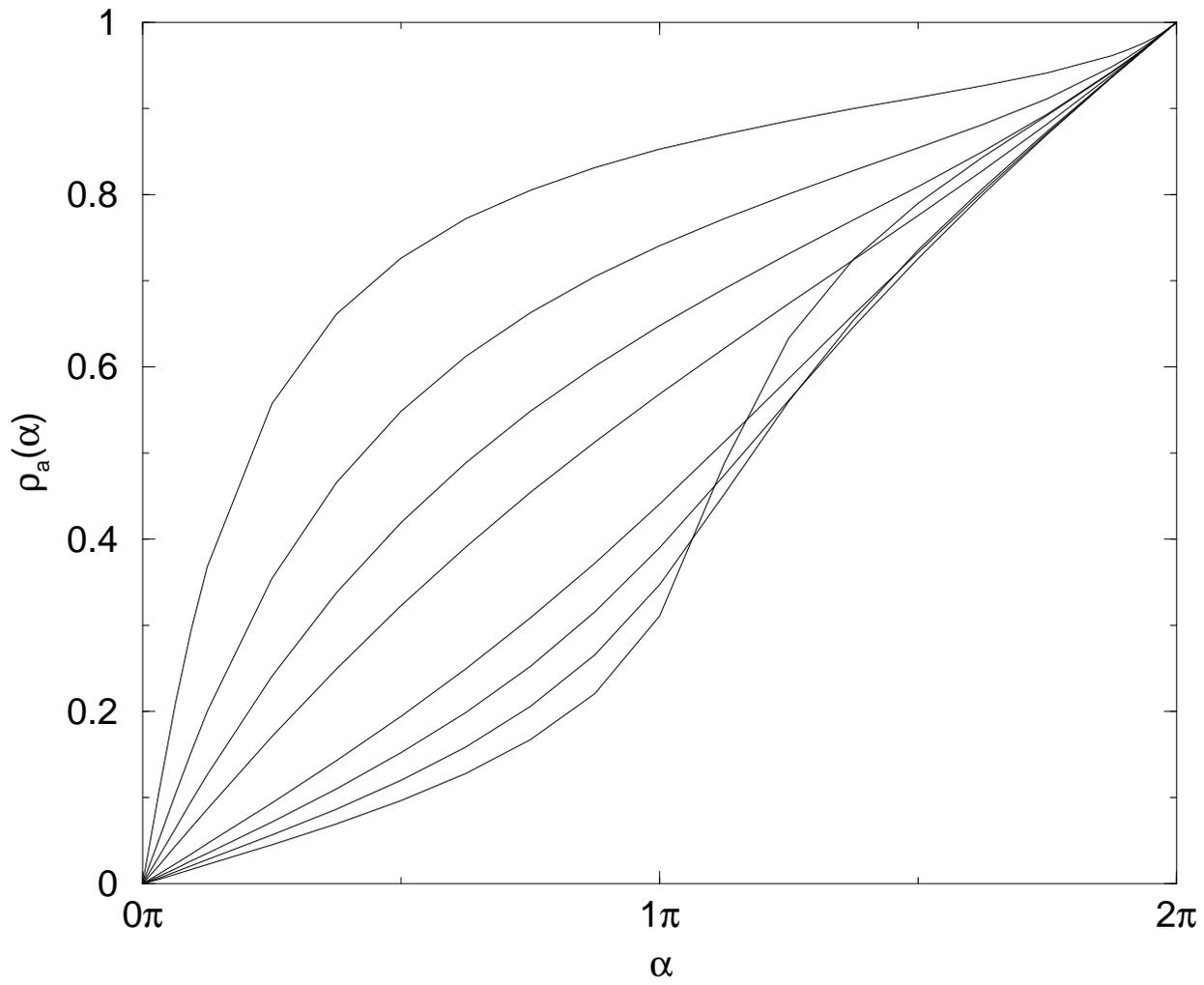}
\caption{Plot  of $\rho_a{(\alpha)}$  against $\alpha$,  $a\ =\  0.8 $
(top curve)$,\, 0.6,\, 0.4,\ 0.2,$ also the alternating cases, $ a\ =\
-0.2,\  -0.4,\ -0.6,\  -0.8$  (bottom curve).   The  lines are  linear
interpolations between the discrete data points.}
\label{f10}
\end{figure}

\begin{figure}
\epsfbox{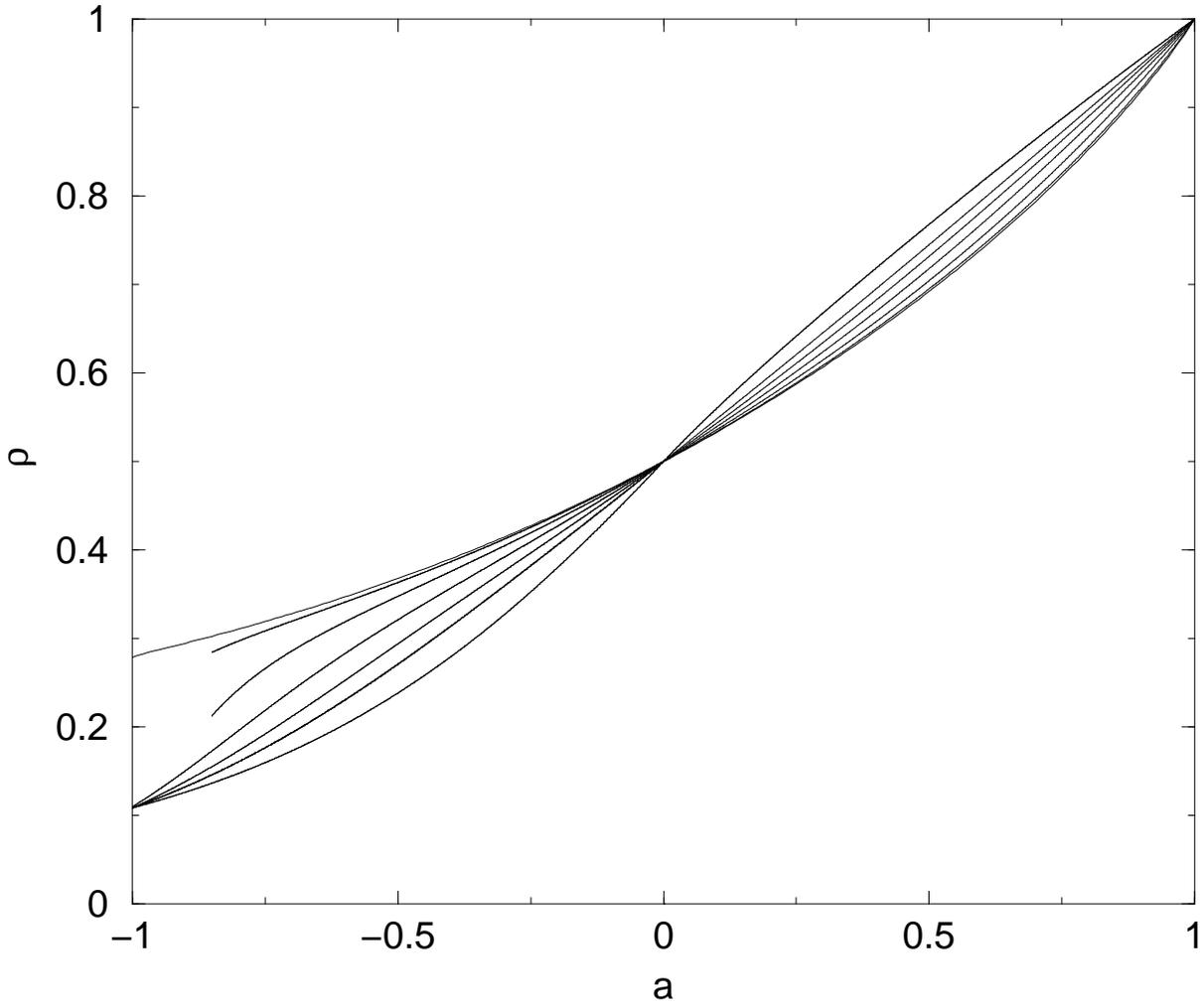 }
\caption{Plot of  $\rho(a)$, $a=e^{-\alpha \Delta  \!T}$, $\alpha=1/2$
in all cases, with  $a<0$ denoting alternating persistence.  For $a>0$
the  curves are,  from  the bottom:  random  walk ($\beta=\infty$)  to
$O(a^{49})$, and constrained Pad\'es $\beta=24$, $\beta=6$, $\beta=3$,
$\beta=2$,  $\beta=3/2$,  $\beta=1$  to  $O(a^{25})$.  For  $a<0$  the
ordering  is reversed,  the  random walk  is  $O(a^{49})$, the  random
accelerations  are  $O(a^{19})$.  The  Pad\'ed  curves  shown are  the
averages of the diagonal Pad\'e  (for odd orders) and the four nearest
off-diagonal Pad\'es, although these do not differ visibly (except for
the plots  of $\theta(a)$ for $a$  larger than about  0.99, see figure
\ref{f25}).  Only Pad\'es which do not have spurious poles on the real
axis  in  the range  $-0.25<a<1.25$  are  considered.   Note that  for
$\beta$  finite, $\rho(-1) \approx  0.108$ while  for the  random walk
$\rho(-1)=0.280 \dots$.   As $\beta \to \infty$, the  turn down occurs
closer  to $a=-1$  and is  sharper, and  so for  $\beta=6$,  $24$, the
Pad\'e can  no longer `predict' the  curves and in these  2 cases only
the raw series are plotted. }
\label{f20}
\end{figure}

\begin{figure}
\epsfbox{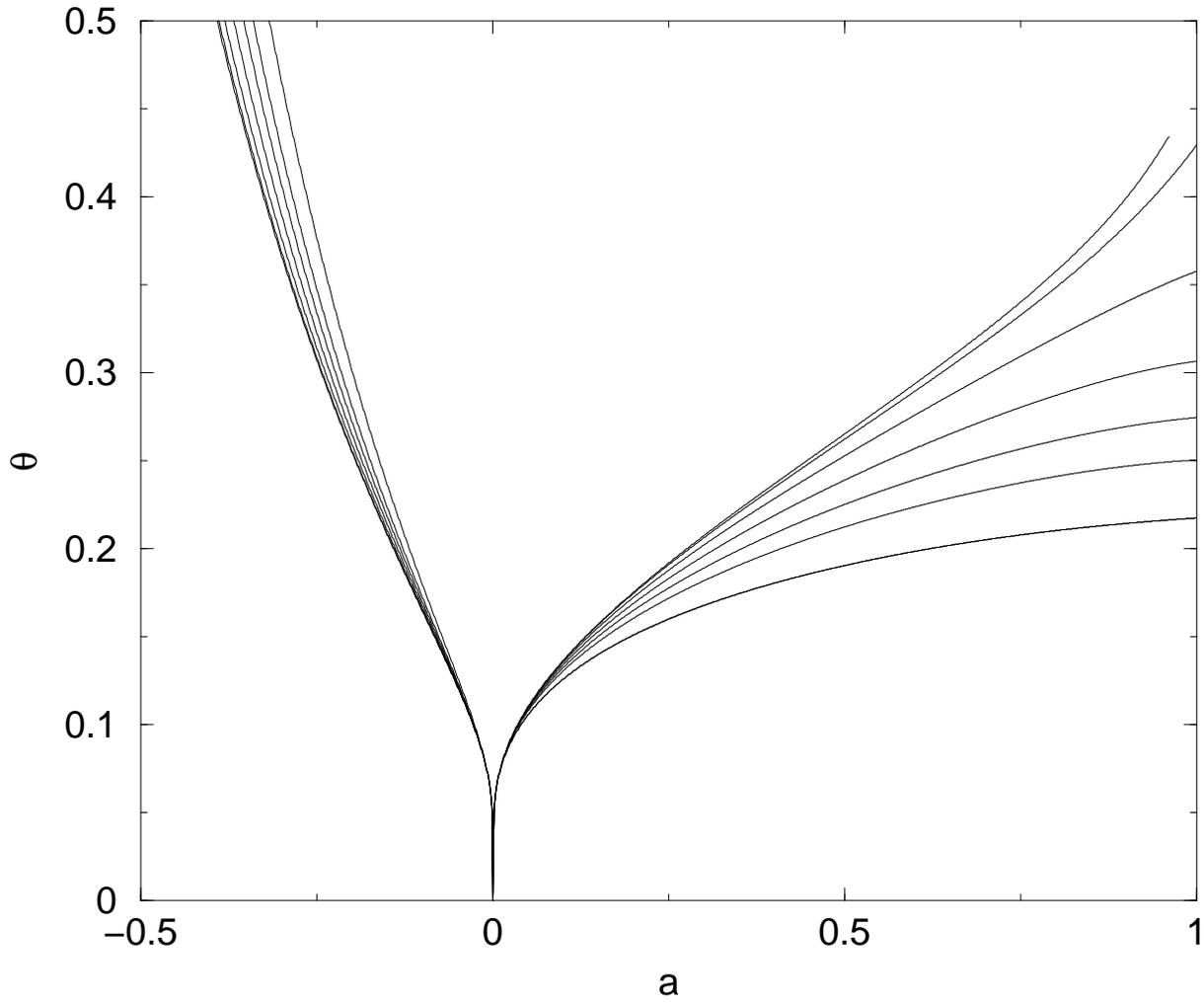}
\caption{Plot     of      $\theta     (a)$     $(={1      \over     2}
\rm{ln}(\rho)/\rm{ln}(a))$,  $a=e^{-\alpha \Delta  \!T}$, $\alpha=1/2$
in all cases, where $\rho$ is the same as for figure \ref{f20}.  $a<0$
denotes alternating  persistence.  For $a>0$ the curves  are, from the
top: random  walk ($\beta=\infty$), $\beta=24$,  $\beta=6$, $\beta=3$,
$\beta=2$, $\beta=3/2$, $\beta=1$.  For  $a<0$, the order is reversed.
Note that the random walk series  has not converged for $a \to 1$, but
is $1/2$ in this limit. }
\label{f25}
\end{figure}

\begin{figure}
\epsfbox{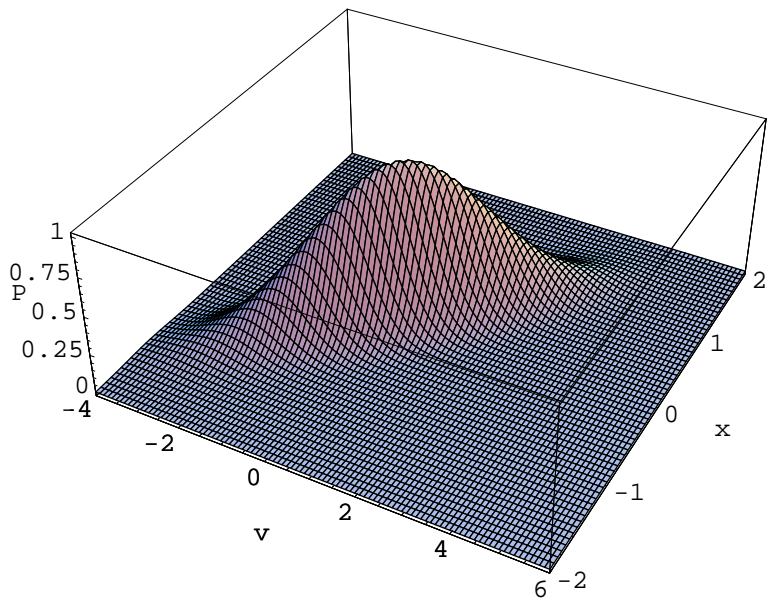}
\caption{The Eigenfunction for $\alpha=1/2$, $\beta=3/2$ and $a=0.1$.}
\label{ddxeigenfn}
\end{figure}

\newpage

\begin{figure}
\epsfbox{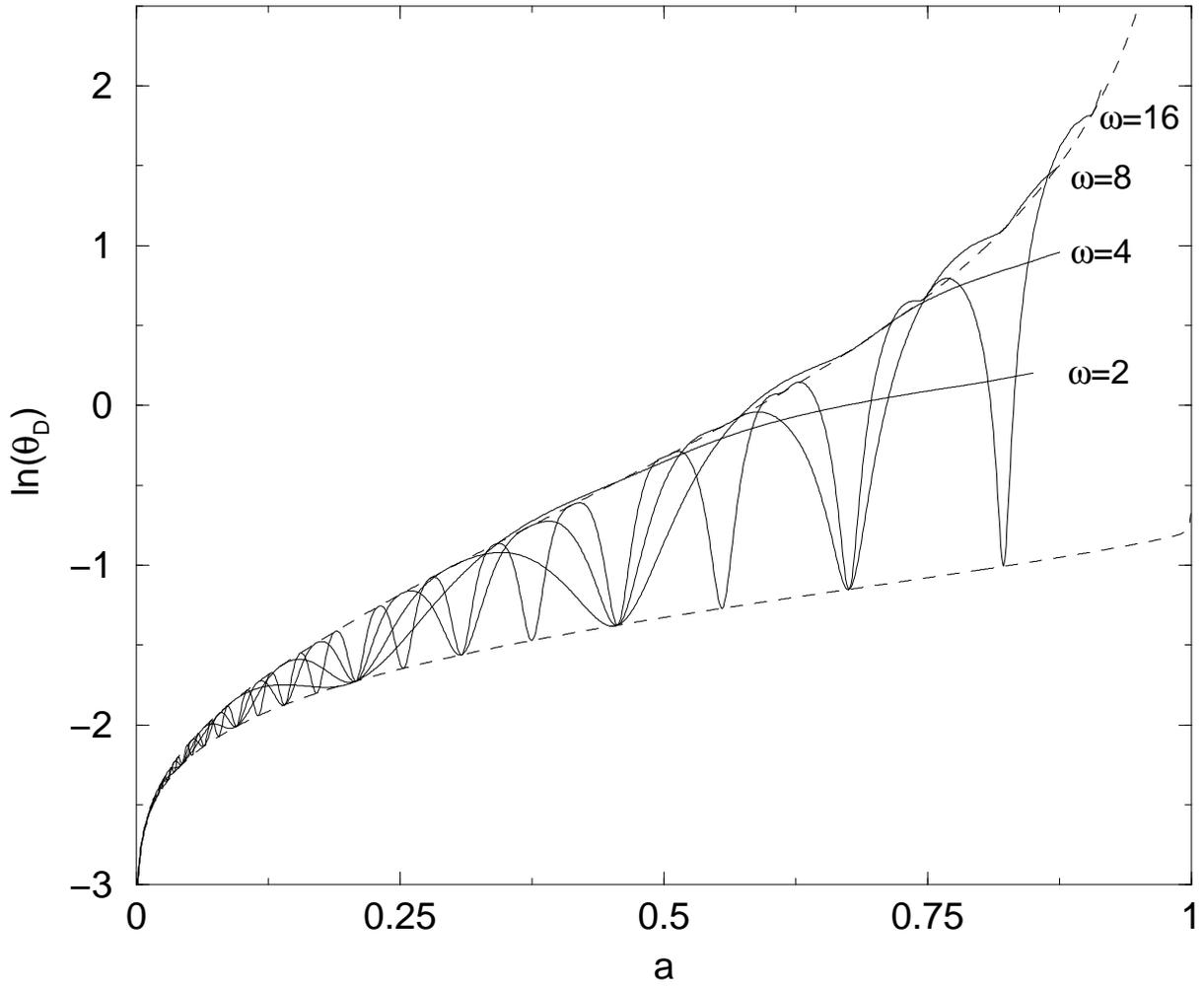}
\caption{Plots  of  the  underdamped  noisy SHO  persistence  exponent
against   $a=e^{-\Delta\!T/2}$,   calculated   using  the   correlator
expansion for  the cases $\gamma=1/2$ with  $\omega=2,4,8,16$ from the
lower right  respectively.  Also plotted are the  discrete random walk
(lowest  dashed curve)  and  alternating discrete  random walk  (upper
dashed curve) which are equal to the oscillating exponents for $\omega
\Delta\!T=2m\pi$  and $\omega  \Delta\!T=(2m+1)\pi$  respectively, $m$
integer.  Note that the series have not converged for $a \to 1$.}
\label{ddxunderdampedtheta}
\end{figure}


\begin{references}

\bibitem{manyrefincmajumdarsdotxtpowalpha} A. Dhar and S. N.  Majumdar,
Phys.\ Rev.\  E {\bf  59}, 6413 (1999);   G. De Smedt,  C.  Godr\`eche,
and J. M. Luck, J. Phys.\ A {\bf 34}, 1247 (2001).

\bibitem{burk} T. W. Burkhardt, J. Phys.\ A {\bf 26}, L1157 (1993).

\bibitem{inelasticcollapseofarandomlyforcedparticle}   S. J. Cornell,
M. R. Swift, and A.J. Bray, Phys.\ Rev.\ Lett. {\bf 81}, 1142 (1998).

\bibitem{inelasticcollapseofarandomlyforcedparticle2} M. R. Swift and
A. J. Bray, Phys.\ Rev.\ E {\bf 59}, R4721 (1999).

\bibitem{IIA1} S. N. Majumdar, C. Sire, A. J. Bray, and S. J. Cornell,
Phys.\ Rev.\ Lett.\ {\bf 77}, 2867 (1996).

\bibitem{IIA2}  B. Derrida, V. Hakim, and R. Zeitak, Phys.\ Rev.\ 
Lett.\ {\bf 77}, 2871 (1996).

\bibitem{DBG} B. Derrida, A. J. Bray,  and C. Godr\`eche, 
J. Phys.\ A {\bf 27},L357 (1994).

\bibitem{derrida1}  P. L. Krapivsky, E. Ben-Naim, and S. Redner,
Phys.\ Rev.\ E {\bf 50}, 2474 (1994).

\bibitem{derrida2}  B. Derrida,  V. Hakim,  and   V. Pasquier,
Phys.\ Rev.\ Lett.\ {\bf 75}, 751  (1995); J. Stat.\ Phys.\ {\bf 85},
763 (1996).

\bibitem{bray} A. J. Bray, B. Derrida, and C. Godr\`eche,
Europhys.\ Lett.\ {\bf 27}, 175 (1994).

\bibitem{satyaandsire2}  C. Sire, S. N. Majumdar, and  A. Rudinger,
Phys.\ Rev.\  E {\bf 61}, 1258 (2000). 

\bibitem{majsire} S. N. Majumdar, and C. Sire, Phys.\ Rev.\  Lett.\ 
{\bf 77}, 1420 (1996).

\bibitem{crit} S. N. Majumdar, A. J. Bray, S. J. Cornell, and C. Sire,
Phys.\ Rev.\ Lett.\ {\bf 77}, 3704 (1996).

\bibitem{clement} S. Cueille and C.  Sire, J. Phys.\  A {\bf 30}, L791
(1997); Euro.\ Phys.\ J. B, {\bf 7}, 111 (1999).

\bibitem{krapiv} P. L. Krapivsky, and E. Ben-Naim, Phys.\ Rev.\  E {\bf
56}, 3788 (1998).

\bibitem{majcor} S. N. Majumdar and S.J.  Cornell, Phys.\ Rev.\ E {\bf
57}, 3757 (1998). 

\bibitem{lee} B. P. Lee and A. D. Rutenberg, Phys.\ Rev.\ Lett.\ 
{\bf 79}, 4842 (1997); A. D. Rutenberg and B. P. Lee, Phys.\ Rev.\ 
Lett.\ {\bf 83}, 3772 (1999). 

\bibitem{krug} J. Krug, H. Kallabis, S. N. Majumdar, S. J. Cornell, 
A. J. Bray, and C. Sire, Phys.\ Rev.\ E {\bf 56}, 2702 (1997).

\bibitem{kallabis} H. Kallabis and J. Krug, Europhys.\ Lett.\ 
{\bf 45}, 20 (1999). 

\bibitem{tnd}  Z. Toroczkai,  T. J. Newman,  and  S. Das  Sarma,
Phys.\ Rev.\ E {\bf 60}, R1115 (1999).

\bibitem{cardy}  J. Cardy, J. Phys.\ A  {\bf  28},  L19  (1995);
E. Ben-Naim, Phys.\ Rev.\ E {\bf 53}, 1566 (1996); M. Howard, 
J. Phys.\ A {\bf 29}, 3437 (1996); C. Monthus, Phys.\ Rev.\ E 
{\bf 54}, 4844 (1996). 

\bibitem{fisher} D. S. Fisher, P. Le Doussal, and C. Monthus, 
Phys.\ Rev.\ Lett.\ {\bf 80}, 3539 (1998); Phys.\ Rev.\ E {\bf 59}, 
4795 (1999); P. Le Doussal and C. Monthus, Phys.\ Rev.\ E {\bf 60}, 
1212 (1999). 

\bibitem{steve1} A. J. Bray and S. J.  O'Donoghue, Phys.\ Rev.\ 
E {\bf 62}, 3366 (2000).

\bibitem{steve2} S. J. O'Donoghue and A. J. Bray, Phys.\ Rev.\ E 
{\bf 64}, 041105 (2001).

\bibitem{review} S. N. Majumdar, Curr.\ Sci.\ {\bf 77}, 370 (1999),
also available on cond-mat/9907407.

\bibitem{marcos} M. Marcos-Martin, D. Beysens, J-P.\ Bouchaud, C.
Godr\`eche, and I. Yekutieli, Physica A {\bf 214}, 396 (1995).

\bibitem{yurke}  B. Yurke,  A. N. Pargellis,  S. N. Majumdar,  and
C. Sire, Phys.\ Rev.\ E {\bf 56}, R40 (1997).

\bibitem{tam} W. Y. Tam, R. Zeitak, K. Y. Szeto, and  J.  Stavans,
Phys.\ Rev.\ Lett.\ {\bf 78}, 1588 (1997).

\bibitem{wong}  G. P.\ Wong, R. W. Mair, R. L. Walsworth, and  
D. G. Cory, Phys.\ Rev.\ Lett.\ {\bf 86}, 4156 (2001). 

\bibitem{MBE} S.N.  Majumdar,  A. J. Bray, and G. C. M. A. Ehrhardt, 
Phys.\ Rev.\ E {\bf 64}, 015101(R) (2001). 

\bibitem{fractalrefs}  M. Kac,  SIAM   Review  {\bf  4},  1  (1962);
I. F. Blake and W. C.  Lindsay, IEEE Trans.\ Information Theory {\bf19},
295 (1973).

\bibitem{MD} S. N. Majumdar and D. Dhar, Phys.\ Rev.\ E {\bf 64}, 
046123 (2001).

\bibitem{M} S. N. Majumdar, preprint, cond-mat/0110221.

\bibitem{ctl} G. C. M. A. Ehrhardt and A. J. Bray, cond-mat/0109526, 
to appear in Phys.\ Rev.\ Lett. 

\bibitem{sinai} Y. G. Sinai, Theor.\ Math.\ Phys.\ {\bf 90}, 219 (1992).

\bibitem{alanetal}  K. Oerding,  S. J. Cornell,  and A. J.  Bray, 
Phys.\ Rev.\ E {\bf 56}, R25 (1997).

\bibitem{zeitakshorttimeexpansion} R. Zeitak, Phys.\ Rev.\ E {\bf 56},
2560 (1997). 

\bibitem{DombAndGreenVol3} D. S. Gaunt and A. J. Guttmann, in {\it Phase  
Transitions and Critical  Phenomena}, Vol.\ 3 (eds.\  C.  Domb and M. S. 
Green), p202-3 (1974). 

\end{references}
\end{document}